\documentclass[longauth]{aa}

\usepackage[dvipsnames]{xcolor}
\usepackage{lscape, threeparttable}
\usepackage{adjustbox}
\usepackage{float}
\usepackage{graphicx}
\usepackage{multirow,xspace,upgreek}
\usepackage{txfonts}
\newcommand{\mic}{\ensuremath{\upmu}m\xspace}
\newcommand{\Teff}{\ensuremath{\mathrm{T_{eff}}}\xspace}
\newcommand{\fsed}{\ensuremath{\mathrm{f_{sed}}}\xspace}

\newcommand{\Rjup}{\ensuremath{\mathrm{R_{Jup}}}\xspace}
\newcommand{\Mjup}{\ensuremath{\mathrm{M_{Jup}}}\xspace}
\newcommand{\Lbol}{\ensuremath{\mathrm{L_{bol}}}\xspace}
\newcommand{\Rlam}{\ensuremath{\mathrm{R_{\uplambda}}}\xspace}

\usepackage{natbib}
\usepackage[colorlinks=true,linkcolor=blue,allcolors=blue]{hyperref}

\begin{document}

  \title{X-SHYNE: X-Shooter spectra of young exoplanet analogs\thanks{Based on observations collected at the European Organization for Astronomical Research in the Southern Hemisphere under ESO programs 0101.C-0290; 0102.C-0121; 0103.C-0231; 0104.C-0094; 111.24PX.001; and 114.27B1.001}}

    \subtitle{II. Presentation and analysis of the full library}

    \author{
        S.~Petrus\thanks{NASA Postdoctoral Program Fellow}\inst{\ref{nasa}, \ref{udp}, \ref{yems}, \ref{ipag}}
        \and
        G.~Chauvin\inst{\ref{heidelberg},\ref{oca}}
        \and
        M.~Bonnefoy\inst{\ref{ipag}}
        \and
        P.~Tremblin\inst{\ref{cea}}
        \and
        C.~Morley\inst{\ref{texas}}
        \and
        B.~Charnay\inst{\ref{lesia}}
        \and
        G.~Suarez\inst{\ref{amnh}}
        \and
        J.~Gagn\'e\inst{\ref{montreal}, \ref{montreal2}}
        \and
        P.~Palma-Bifani\inst{\ref{oca}, \ref{lesia}}
        \and
        A.~Denis\inst{\ref{lam}}
        \and
        M.~Ravet\inst{\ref{ipag}}
        \and 
        A.~Bayo\inst{\ref{munich}}
        \and
        B.~Bézard\inst{\ref{lesia}}
        \and
        B.~Biller\inst{\ref{Edinburgh}}
        \and
        P.~Delorme\inst{\ref{ipag}}
        \and
        J.~Faherty\inst{\ref{amnh}}
        \and
        J.-M.~Goyal\inst{\ref{cornel}, \ref{Exeter}}
        \and
        K.~Hoch\inst{\ref{stsci}}
        \and
        K.~Hoy\inst{\ref{udp}, \ref{yems}}
        \and
        J.-S.~Jenkins\inst{\ref{udp}, \ref{Casilla}}
        \and
        A.-M.~Lagrange \inst{\ref{lesia}}
        \and
        B.~Lavie\inst{\ref{geneve}}
        \and
        M.~C.~Liu\inst{\ref{hawaii}}
        \and
        E.~Manjavacas\inst{\ref{stsci}, \ref{hopkins}}
        \and
        G.-D.~Marleau\inst{\ref{bernWP}, \ref{heidelberg}, \ref{duisburg}}
        \and 
        M.~McElwain \inst{\ref{nasa}}
        \and 
        P.~Molli\`ere \inst{\ref{heidelberg}}
        \and
        C.~Mordasini\inst{\ref{bernWP}, \ref{bernCSH}}
        \and
        M.~Phillips\inst{\ref{hawaii}}
        \and
        P.~Rojo\inst{\ref{lascondes}}
        \and
        Z.~Zhang\thanks{NASA Sagan Fellow}\inst{\ref{santacruz}, \ref{rochester}}
        \and
        A.~Zurlo\inst{\ref{udp}, \ref{yems}}
        }
        
   \institute{
    NASA-Goddard Space Flight Center, Greenbelt, MD 20771, USA\\ \email{simon.petrus.pro@gmail.com}
    \label{nasa}
   \and
    Instituto de Estudios Astrofísicos, Facultad de Ingeniería y Ciencias, Uni.\ Diego Portales, Av. Ejército 441, Santiago, Chile
    \label{udp}
   \and
   Millennium Nucleus on Young Exoplanets and their Moons (YEMS), Santiago, Chile
    \label{yems}
    \and
   Univ. Grenoble Alpes, CNRS, IPAG, F-38000 Grenoble, France
    \label{ipag}
   \and    
    Max-Planck-Institut f\"{u}r Astronomie, K\"{o}nigstuhl 17, 69117 Heidelberg, Germany
    \label{heidelberg}
   \and
   Laboratoire Lagrange, Université Cote d’Azur, CNRS, Observatoire de la Cote d’Azur, 06304 Nice, France
    \label{oca}
   \and
   Maison de la Simulation, CEA, CNRS, Univ. Paris-Sud, UVSQ, Université Paris-Saclay, 91191 Gif-sur-Yvette, France
    \label{cea}
    \and
    Department of Astronomy, University of Texas at Austin, Austin, TX, USA
    \label{texas}    
    \and
    LESIA, Observatoire de Paris, Universit\'{e} PSL, CNRS, Sorbonne Universit\'{e}, Univ. Paris Diderot, Sorbonne Paris Cit\'e, 5 place Jules Janssen, 92195 Meudon, France
    \label{lesia}
    \and
    Department of Astrophysics, American Museum of Natural History, Central Park West at 79th Street, NY 10024, USA
    \label{amnh}
    \and
    Plan\'etarium de Montr\'eal, Espace pour la Vie, 4801 av. Pierre-de Coubertin, Montr\'eal, Qu\'ebec, Canada
    \label{montreal}
    \and
    Trottier Institute for Research on Exoplanets, Universit\'e de Montr\'eal, D\'epartement de Physique, C.P.~6128 Succ. Centre-ville, Montr\'eal, QC H3C~3J7, Canada
    \label{montreal2}
    \and
    Aix Marseille Univ, CNRS, CNES, LAM, Marseille, France
    \label{lam}
    \and
    European Southern Observatory, Karl Schwarzschild-Stra\ss e 2, D-85748 Garching bei M\"{u}nchen, Germany
    \label{munich}
    \and
    Institute for Astronomy The University of Edinburgh Royal Observatory Blackford Hill Edinburgh EH9 3HJ UK
    \label{Edinburgh}
    \and
    Department of Astronomy and Carl Sagan Institute, Cornell University, 122 Sciences Drive, Ithaca, NY, 14853, USA
    \label{cornel}
    \and
    Astrophysics Group, School of Physics and Astronomy, University of Exeter, Exeter EX4 4QL, UK
    \label{Exeter}
    \and
    AURA for the European Space Agency (ESA), ESA Office, Space Telescope Science Institute, 3700 San Martin Drive, Baltimore, MD, 21218 USA
    \label{stsci}
    \and
    Centro de Astrofísica y Tecnologías Afines (CATA), Casilla 36-D, Santiago, Chile
    \label{Casilla}
    \and
    Observatoire de Genève, Département d’Astronomie, Université de Genève, Chemin Pegasi 51b, 1290 Versoix, Switzerland
    \label{geneve}    
    \and
    Institute for Astronomy, University of Hawai’i at Mānoa, 2680 Woodlawn Drive, Honolulu, HI 96822, USA
    \label{hawaii}
    \and
    Department of Physics \& Astronomy, Johns Hopkins University, Baltimore, MD 21218, USA
    \label{hopkins}
    \and
    Division of Space Research and Planetary Sciences, Physics Institute, University of Bern, Gesellschaftsstr.~6, 3012 Bern, Switzerland
    \label{bernWP}
    \and
    Center for Space and Habitability, University of Bern, Gesellschaftsstr.~6, 3012 Bern, Switzerland
    \label{bernCSH}
    \and
    Fakult\"at f\"ur Physik, Universit\"at Duisburg-Essen, Lotharstra\ss{}e 1, 47057 Duisburg, Germany
    \label{duisburg}    
    \and
    Departamento de Astronomía, Universidad de Chile, Camino el Observatorio 1515, Las Condes, Santiago, Chile
    \label{lascondes}        %
    \and
    Department of Astronomy \& Astrophysics, University of California, Santa Cruz, CA 95064, USA
    \label{santacruz}
    \and
     Department of Physics \& Astronomy, University of Rochester, Rochester, NY 14627, USA
    \label{rochester}
   }
   \date{Received 01/03/25; accepted 05/30/25}

  \abstract{The characterization of exoplanetary spectra is a crucial step in understanding the chemical and physical processes shaping their atmospheres and constraining their formation and evolutionary history. The X-SHYNE library is a homogeneous sample of 43 medium-resolution (\Rlam~$\sim$~8000) infrared (0.3–2.5~\mic) spectra of young ($<$500~Myr), low-mass ($<$20~\Mjup), and cold (\Teff~$\sim$600–2000~K) isolated brown dwarfs and wide-separation companions observed with the VLT/X-Shooter instrument. To characterize our targets, we performed a global comparative analysis. We first applied a semi-empirical approach. By refining their age and bolometric luminosity, we derived key atmospheric and physical properties, such as \Teff, mass, surface gravity (g), and radius, using the evolutionary model \texttt{COND03}. These results were then compared with the results from a synthetic analysis based on three self-consistent atmospheric models: the cloudy models \texttt{Exo-REM} and \texttt{Sonora Diamondback}, and the cloudless model \texttt{ATMO}. To compare our spectra with these grids we used the Bayesian inference code \texttt{ForMoSA}. We found similar \Lbol estimates between both approaches, but an underestimated \Teff from the cloudy models, likely due to a lack of absorbers that could dominate the J and H bands of early L. We also observed a discrepancy in the log(g) estimates, which are dispersed between 3.5 and 5.5 dex for mid-L objects. We interpreted this as a bias caused by a range of rotational velocities leading to cloud migration toward equatorial latitudes, combined with a variety of viewing angles that result in different observed atmospheric properties (cloud column densities, atmospheric pressures, etc.). This interpretation is supported by the correlation of the color anomaly $\Delta$(J-K) of each object with log(g) and the parameter \fsed that drives the sedimentation of the clouds. Finally, while providing robust estimates of [M/H] and C/O for individual objects remains challenging, the X-SHYNE library globally suggests solar values, which are consistent with a formation via stellar formation mechanisms. This study highlights the strength of homogeneous datasets in performing comparative analyses, reducing the impact of systematics, and ensuring robust conclusions while avoiding over-interpretation.}  


       \maketitle

\section{Introduction}
\label{sec:introduction}
For the past twenty years, it has been possible to obtain spectra of massive planetary-mass companions ($>$ 3~\Mjup) at wide separations ($>$ 15~AU) observed through direct imaging mostly in near-infrared. Initially, this was achieved from the ground at low resolution with high-contrast spectro-imagers like VLT/SPHERE (\Rlam~$\sim$~50; \citealt{Beuzit19}), Keck/GPI (\Rlam~$\sim$~66; \citealt{Macintosh14}), and Subaru/SCExAO+CHARIS (\Rlam~$\sim$~80; \citealt{Jovanovic16}, \citealt{Peters12}), at medium resolution with integral field spectrographs such as Keck/OSIRIS (\Rlam~$\sim$~8000; \citealt{Larkin06}), VLT/SINFONI (\Rlam~$\sim$~4000; \citealt{Eisenhauer03}), or now VLT/ERIS (\Rlam~$\sim$~8000; \citealt{Davies23}), and with interferometry using VLTI/GRAVITY (\Rlam~$\sim$~500-4000; \citealt{GRAVITY17}) and VLTI/MATISSE (\Rlam~$\sim$~500; \citealt{2022A&A...659A.192L}), and finally at high resolution with projects like VLT/CRIRES+ (\Rlam~$\sim$~100,000; \citealt{2004SPIE.5492.1218K}), Keck/KPIC (\Rlam~$\sim$~35,000; \citealt{Mawet17}), VLT/HiRISE (\Rlam~$\sim$~100,000; \citealt{Vigan22}), and Subaru/REACH (\Rlam~$\sim$~100,000; \citealt{Kotani20}). More recently, JWST has enriched this instrumental landscape from space by offering a combination of integral field spectroscopy and medium spectral resolution up to 28~\mic with its instruments NIRSpec (\Rlam~$\sim$~2700; \citealt{Boker22}) and MIRI (\Rlam~$\sim$~3500; \citealt{Wells15}) that can image companions at large separation ($>$100 mas). Naturally, the spectroscopic information contained in these data is exploited to characterize the atmosphere and the dynamic properties of the observed sources, with the spectral resolution guiding the type of parameters accessible.

With low-resolution spectra coupled with photometric data, it has been possible to estimate the effective temperature \Teff and surface gravity log(g) of observed objects by analyzing the modified blackbody spectrum with significant water absorptions (e.g., 51~Eri~b \citealt{Macintosh15}; HD~95086~b \citealt{Desgrange22}; AF~Lep~b \citealt{Palma24}). Medium resolution has enabled the detection of molecules and atoms, enabling the estimation of relative chemical compositions such as the metallicity ([M/H]) and carbon-oxygen ratio (C/O) (e.g., HR~8799~c \citealt{Konopacky13}; VHS~1256~b \citealt{Hoch22}; HIP~65426~b \citealt{Petrus21}; AB~Pic~b \citealt{Palma23}) and isotopic ratios (e.g., TYC~8998-760-1~b \citealt{Zhang21}). Finally, the high resolution provides access to the vertical structure of planetary atmospheres as well as dynamic properties of companions, such as projected rotational velocity v~sin(i) and radial velocity (e.g., $\beta$~Pic~b \citealt{Snellen14}; system HR~8799 \citealt{Wang21}).

Estimating these atmospheric properties is now a critical challenge in exoplanetary science because they are potentially linked to the formation and evolution mechanisms of planetary systems. The [M/H] of atmospheres is considered a possible tracer of the amount of solid material accreted during formation \citep{Ormel21}, while the C/O or the carbon monoxide isotopologue ratio $^{13}$CO/$^{12}$CO can provide insights into the formation distance of planets relative to their star, particularly, relative to the location of the ice lines of key chemical elements like H$_{2}$O, CO$_{2}$, and CO \citep{Oberg11}. Although \cite{Molliere22} highlighted the interpretive limitations of these tracers, especially within the context of dynamic chemistry in disks, they remain crucial properties for constraining the structure and evolution of planetary atmospheres.

One approach used to constrain these properties is to compare the observed spectra of planetary companions with synthetic spectra generated from atmospheric model predictions. The current versions of the different families of models available allow for the generation of synthetic spectra at medium and high spectral resolution over a wide wavelength coverage. Thus, it is possible to fit the observed data using their full potential, i.e., by maximizing both the spectral resolution and the wavelength range. However, due to the computational time required for such inversions, this method is mainly applicable using precompiled grids of synthetic spectra generated from self-consistent atmospheric models, which fix part of the physics and chemistry of the atmospheres with strong assumptions, thus limiting the number of explored parameters (less than 5) but reducing the computational time required for fitting. Indeed, with this approach, it is sufficient to interpolate within the grid, unlike retrieval methods that parametrize the physics and the chemistry of the atmospheres and calculate a new spectrum at each iteration based on the considered physics. Generally, reduced resolution spectra are used in this second case (\citealt{Zhang25}, \citealt{Matthews25}, \citealt{Whiteford25}). The work presented here focuses on the use of precompiled grids.

Over the past fifteen years, various grids of synthetic spectra have been developed and used. Each of these grids is generated by atmospheric models that differ in complexity (cloud formation and evolution, non-equilibrium chemistry, micro-physics, initial chemical abundances, etc.), parameter space explored, and wavelength coverage and resolution. A description of this diversity of models is detailed in \cite{Petrus24}, who used them to fit medium-resolution spectra of the companion VHS~1256~b (L7) obtained with JWST/NIRSpec and JWST/MIRI \citep{Miles23}. In this study, \cite{Petrus24} used the code \texttt{ForMoSA} \citep{Petrus20, Petrus23} to demonstrate the limitations of these models by showing that they failed to represent data over an extended wavelength range. One of the main reasons cited to explain these systematic errors is the complexity of the cloud properties present in the atmosphere of this object. To account for these systematic errors, they proposed a fitting strategy based on the definition of reduced spectral windows, fitted independently. Although this technique significantly improves the quality of the fits, the robust estimation of C/O and [M/H] remains challenging. Indeed, the estimation of parameters depends on the spectral window as well as on the model considered. While the results of this work confirm the conclusions of previous studies using self-consistent atmospheric models \citep{Petrus20, Palma23}, they are only valid for VHS~1256~b, known to be subject to significant photometric variability \citep{Bowler20, Zhou20, Zhou22}, likely due to a complex atmospheric structure and inhomogeneous cloud coverage, as expected for an object at the L-T transition \citep{Burgasser02, Cushing08, Marley10}. The impact of these systematic errors on other spectral type ranges remains to be explored.

Today, the number of exoplanetary companions for which we have medium-resolution spectra over an extended wavelength range is limited due to the glare of their host stars. This prevents a homogeneous atmospheric characterization study over a wide temperature range. However, this diversity can be found in the population of young, isolated, low-mass brown dwarfs. Young, nearby associations have been identified to contain isolated very-low-mass stars, sub-stellar objects, and planetary mass objects.
Some members of these associations overlap in spectral type, effective temperature, age, and mass with currently known directly imaged exoplanets. These similar physical properties are expected to imply similar atmospheric properties, and therefore, this population of young isolated planetary-mass objects constitutes an ideal laboratory to explore physical processes at play in the atmospheres of imaged exoplanets. 

However, for these isolated objects, a star-like formation through the gravitational collapse of molecular clouds is a plausible scenario. Therefore, despite their similarities with exoplanets, subtle differences in their atmospheric chemical compositions are expected. Consequently, characterizing their atmospheres offers a valuable point of comparison for exploring alternative formation pathways proposed for exoplanets, such as core accretion and gravitational instability.

Important efforts have been devoted to acquiring medium- and high-resolution spectra of these objects without host star in a wide range of spectral types (early-M to late-L) and a large wavelength coverage (0.5 to 2.5 $\upmu$m), to form libraries of planetary analogs spectra \citep{Bonnefoy14, Gagne2015, Cruz18, Almendros22, Hurt24}. These data can be used as comparison tools for the spectra of imaged companions \citep{Samland17, Chauvin17, Mesa23} to estimate their spectral type, among other things. They also serve as a database for conducting empirical analyses \citep{Allers13, Filippazzo15, Piscarreta24}, primarily based on the calculation of spectral indices or equivalent widths of detected atomic and molecular absorptions. Furthermore, they can act as templates for generating simulated data used in the development of new instruments. Being devoid of a host star, it is also easier to obtain their mid-infrared spectra using space telescopes such as Spitzer or now JWST \citep{Suarez22}. 

In this article, we present the X-SHYNE\footnote{X-Shooter spectra of YouNg Exoplanet analogs.} library, which comprises 43 spectra of brown dwarfs that have been selected to cover a large ranges of age (<500~Myr), masses (<20~\Mjup), and \Teff (600–2000~K), overlapping the fundamental properties of the known imaged companions. X-SHYNE is therefore a library of exoplanet analogs. The data have been obtained with the VLT/X-Shooter instrument to explore an extended wavelength range (0.3–2.5~\mic) at medium-resolution (\Rlam~$\sim$~8000). This unique library allows for probing the information carried by the pseudo-continuum sculpted by effective temperature, clouds, and H$_{2}$-H$_{2}$ and H$_{2}$-He collision-induced absorption (CIA). Additionally, the medium-resolution information can be exploited to estimate chemical composition and thus formation tracers. A first showcase publication targeted VHS\,1256\,b to illustrate the richness of the X-SHYNE library and prepare the analysis of Early Release Science data of JWST on this object \citep{Petrus23}. With this work, we now analyze the complete X-SHYNE library, with three different grids of atmospheric models to characterize their atmospheres and identify the current limits of the models. Section \ref{sec:spec_pres} presents the sample of spectra. Section \ref{sec:semiempirical_analysis} is devoted to our empirical and homogeneous analysis of this library. Section \ref{sec:syntetic_analysis} presents the inversion of our spectra using the three grids of models considered. A discussion of the results is made in Section \ref{sec:discussion}, followed by a general conclusion in Section \ref{sec:conclusion}.

\section{Observations and reduced spectra}
\label{sec:spec_pres}

\begin{figure*}[h!]
\centering
\includegraphics[width=0.9\hsize]{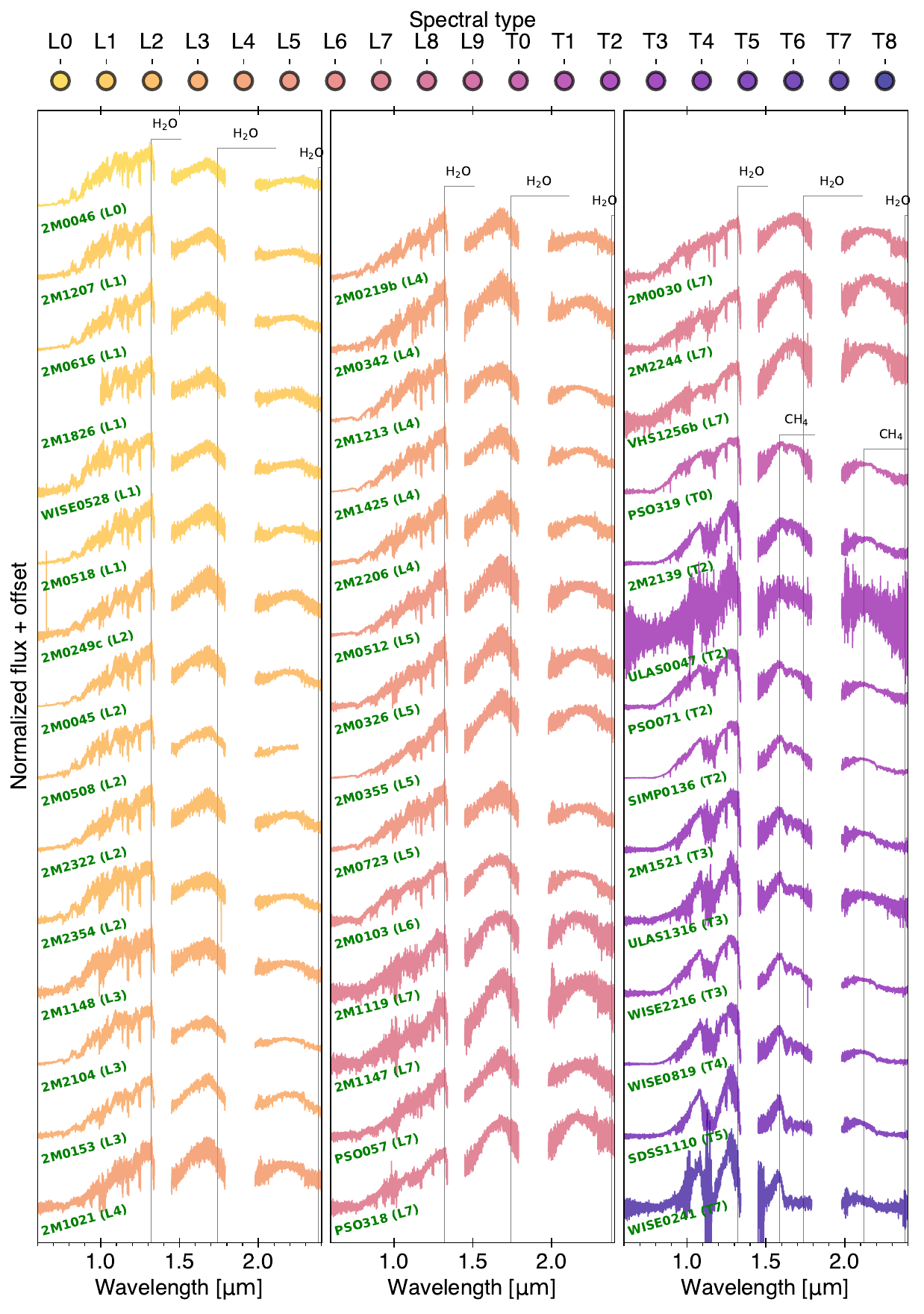}
    \caption{Sequence of the X-SHYNE spectra. The integrated flux between 1.2 and 1.3 \mic normalized the spectra, and an offset was applied. The color of each spectrum indicates its spectral type. Several zooms are given in Figures \ref{fig:sequence_VIS_wav}, \ref{fig:sequence_Y_wav}, \ref{fig:sequence_J_wav}, \ref{fig:sequence_H_wav}, and \ref{fig:sequence_K_wav}.}
    \label{fig:sequence_full_wav}
\end{figure*}

\begin{figure}[h!]
\centering
\includegraphics[width=1.\hsize]{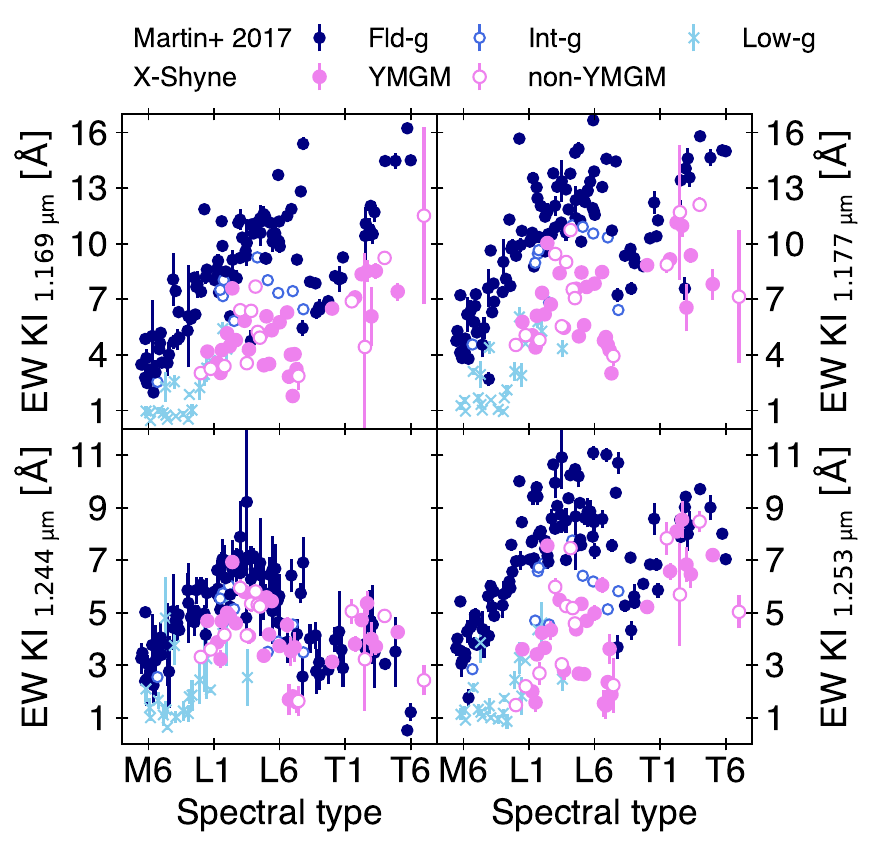}
    \caption{The equivalent widths (EW) of the four potassium absorption lines detected at 1.169, 1.177, 1.244, and 1.253~\mic. To confirm the young age of X-SHYNE's objects, we compared their EWs (pink) with the EW calculated for the low, intermediate, and field gravity objects from the library of spectra published by \cite{Martin17} (blue). To improve the clarity of the plot, objects that share the same spectral type have been evenly distributed within a range corresponding to their spectral type $\pm$1.}
    \label{fig:ki_age}
\end{figure}

The spectra were observed between April 28, 2018, and June 6, 2023, during six different observation periods. The VLT/X-Shooter instrument \citep{Vernet11} was used to obtain medium-resolution spectra (\Rlam$_{UVB}$~$\sim$~3300 for $\lambda \in$ [0.3-0.55]~\mic, \Rlam$_{VIS}$~$\sim$6700 for $\lambda \in$ [0.55-0.1]~\mic, \Rlam$_{NIR}$$\sim$8100 for $\lambda \in$ [0.1-2.5]~\mic) over a wide wavelength coverage (0.3-2.5\mic). The observation dates and the weather conditions during the acquisition are detailed in Table \ref{Tab:obs_log}. The data were reduced using the X-Shooter pipeline version 2.9.3 \citep{Modigliani10} within the ESO \texttt{reflex} environment \citep{Freudling13} to obtain two-dimensional, curvature-corrected, and flux-calibrated traces. The spectrum extraction from these traces was performed with a custom extraction routine (see \citealt{Petrus23} for more details) coupled with the \texttt{molecfit} package \citep{Smette15, Kausch15} for telluric correction.

Figure \ref{fig:sequence_full_wav} shows an overview of the X-SHYNE spectral library after the extraction. As illustrated in the Figures \ref{fig:sequence_VIS_wav}, \ref{fig:sequence_Y_wav}, \ref{fig:sequence_J_wav}, \ref{fig:sequence_H_wav}, and \ref{fig:sequence_K_wav}, a huge diversity of atomic and molecular absorptions can be detected. Thanks to the large range of spectral types explored by X-SHYNE, it is possible to follow the evolution of these features with \Teff. In the optical, the TiO ($\sim$0.69, 0.73, 0.77, 0.84, 0.90, 1.10~\mic) and the VO ($\sim$0.75, 0.80, 1.07~\mic) condensate progressively into solid grains and disappear around the L5 type. This depletion results in an increase in the absorption depth of other elements such as Rb~I (0.795 and 0.780~\mic) and Cs~I (0.894 and 0.852~\mic). For some of our targets, we detect a faint Li~I absorption (0.671~\mic), and for 2MASS~0249~c and 2MASS~0045, we detect an~H$_{\alpha}$ emission line. In the near-infrared, the main atomic features are the K~I doublets (1.177 and 1.169~\mic and 1.252 and 1.243~\mic) which become deeper with decreasing \Teff, and the Na~I doublet (1.140 and 1.138~\mic) which disappears after the L-T transition. We also detect FeH (0.995~\mic) and small traces of CrH (0.8611~\mic). The H-band is sculpted by large absorption bands of H$_{2}$O. Due to the young age of our targets, this band displays a typical triangular shape \citep{Lucas01}, which is completely remodeled after the L-T transition with the appearance of strong absorption bands of CH$_{4}$ long-ward of 1.59~\mic. Lastly, in the K-band, numerous CO overtones are detected from 2.9 to 2.4~\mic, along with a strong absorption band of CH$_{4}$ after the L-T transition.

\begin{table*}[t]
\centering
\caption{The X-SHYNE sample}
\label{Tab:sample_ymg}
\small
\begin{tabular}{ll|ll|l|ll|l}
\hline
\multicolumn{2}{c|}{Target}  &  R.A.         & DEC.          & Sp.T. & Assoc. (Memb. Prob.)   &  Dist.                & Ref.      \\
Simbad & Short           & [hh:mm:ss.ss] & [dd:mm:ss.ss] &       &                        & [pc]                  &           \\ 
\hline
2MASSW J0030300-145033      & 2MASS~0030    & 00:30:30.13   & -14:50:33.40  & L7    & non-YMGM               & 24.3~$\pm$~3.6$^{p}$  & 1, 2      \\
2MASS J00452143+1634446     & 2MASS~0045    & 00:45:21.42   & +16:34:44.74  & L2    & ARG (99.3\%)           & 15.3~$\pm$~0.1$^{p}$  & 3, 4      \\
2MASS J00464841+0715177     & 2MASS~0046    & 00:46:48.42   & +07:15:17.76  & L0    & non-YMGM               & 37.7~$\pm$~0.4$^{p}$  &  3, 4     \\
2MASSI J0103320+193536      & 2MASS~0103    & 01:03:32.04   & +19:35:36.18  & L6    & CRIUS203 (99.6\%) 	    & 27.8~$\pm$~5.2$^{p}$  &  1, 5     \\
2MASS J01531463-6744181     & 2MASS~0153    & 01:53:14.63   & -67:44:18.20  & L3    & Tuc-Hor (91.4\%)	    & 47.0~$\pm$~3.2        & 6         \\
2MASS J02192210-3925225b    & 2MASS~0219~b  & 02:19:22.11   & -39:25:22.54  & L4    & Tuc-Hor (99.7\%)	    & 40.2~$\pm$~0.2$^{p}$  & 7, 4      \\
2MASS J02495436-0558015     & 2MASS~0249~c  & 02:49:54.37   & -05:58:01.58  & L2    & non-YMGM     	        & 49.8~$\pm$~10.5$^{p}$ & 8         \\
2MASS J03264225-2102057     & 2MASS~0326    & 03:26:42.26   & -21:02:05.77  & L5    & ABDMG (95.6\%)	        & 41~$\pm$~4            & 9, 10     \\
2MASSI J0342162-681732      & 2MASS~0342    & 03:42:16.21   & -68:17:32.12  & L4    & Tuc-Hor (95.0\%)	    & 81~$\pm$~1            & 11, 10    \\
2MASS J03552337+1133437     & 2MASS~0355    & 03:55:23.37   & +11:33:43.80  & L5    & ABDMG (97.8\%)	        & 9.2~$\pm$~0.1$^{p}$   & 12, 4     \\
2MASS J05081657-1413479     & 2MASS~0508    & 05:08:16.58   & -14:13:48.04  & L2    & ABDMG	(98.6\%)        & 42.6~$\pm$~1.1$^{p}$  & 13, 4     \\
2MASSI J0512063-294954      & 2MASS~0512    & 05:12:06.37   & -29:49:54.01  & L5    & non-YMGM               & 20.2~$\pm$~1.2$^{p}$  & 14, 15    \\
2MASSI J0518461-275645      & 2MASS~0518    & 05:18:46.17   & -27:56:45.71  & L1    & COL (99.9\%)	        & 54.6~$\pm$~1.9$^{p}$  & 11, 4     \\
2MASS J06165623-2543557     & 2MASS~0616    & 06:16:56.23   & -25:43:55.75  & L1    & non-YMGM   	        & 50.3~$\pm$~24.4$^{p}$ & 16, 17    \\
2MASS J07235265-3309446     & 2MASS~0723    & 07:23:52.66   & -33:09:44.54  & L5    & CARN (99.6\%)          & 32.5~$\pm$~2.3$^{p}$  & 18, 4     \\
2MASS J10212570-2830427     & 2MASS~1021    & 10:21:25.71   & -28:30:42.76  & L4    & non-YMGM               & 43.4~$\pm$~6.4        & 6, 19     \\
TWA 42                      & 2MASS~1119    & 11:19:32.54   & -11:37:46.70  & L7    & non-YMGM   	        &  26.6~$\pm$~6.9       & 20, 21    \\
TWA 41                      & 2MASS~1147    & 11:47:24.21   & -20:40:20.44  & L7    & non-YMGM   	        & 37.5~$\pm$~3.0$^{p}$  & 22, 2     \\
2MASS J11480096-2836488     & 2MASS~1148    & 11:48:00.96   & -28:36:48.90  & L3    & non-YMGM               & 47.8~$\pm$~5.6        & 6         \\
TWA 40                      & 2MASS~1207    & 12:07:48.35   & -39:00:04.48  & L1    & TWA (98.0\%)	        & 65.6~$\pm$~4.4$^{p}$  & 23, 4     \\
2MASS J12130336-0432437     & 2MASS~1213    & 12:13:03.34   & -04:32:43.75  & L4    & CARN (99.3\%)	        & 16.9~$\pm$~0.2$^{p}$  & 14, 4     \\
2MASS J14252798-3650229     & 2MASS~1425    & 14:25:27.98   & -36:50:23.25  & L4    & ABDMG (99.1\%)	        & 11.8~$\pm$~0.1$^{p}$  & 24, 4     \\
2MASS J15210327+0131426     & 2MASS~1521    & 15:21:03.27   & +01:31:42.69  & T3    & non-YMGM   	        & 23.1~$\pm$~3.9$^{p}$  & 25, 2     \\
2MASS J18264679-4602234     & 2MASS~1826    & 18:26:46.80   & -46:02:23.64  & L1    & BPMG (96.0\%)          & 58.1~$\pm$~7.2$^{p}$  & 13, 4     \\
2MASS J21043128-0939217     & 2MASS~2104    & 21:04:31.29   & -09:39:21.82  & L3    & BPMG (83.5\%)          & 53.5~$\pm$~5.0$^{p}$  & 13, 4     \\
CFBDS J213926+022023        & 2MASS~2139    & 21:39:26.77   & +02:20:22.70  & T2    & CRIUS203 (96.6\%)	    & 9.9~$\pm$~0.2$^{p}$   & 26, 27    \\
2MASSW J2206450-421721      & 2MASS~2206    & 22:06:45.00   & -42:17:21.14  & L4    & ABDMG (99.5\%)	        & 29.3~$\pm$~1.2$^{p}$  & 1, 4      \\
2MASS J22443167+2043433     & 2MASS~2244    & 22:44:31.67   & +20:43:43.30  & L7    & ABDMG (98.4\%)	        & 17.0~$\pm$~0.3$^{p}$  & 9, 28     \\
2MASS J23225299-6151275     & 2MASS~2322    & 23:22:53.01   & -61:51:27.53  & L2    & Tuc-Hor (96.0\%)	    & 42.9~$\pm$~1.9$^{p}$  & 29, 4     \\
PSO J358.5527+22.1393       & 2MASS~2354    & 23:54:12.57   & +22:08:22.58  & L2    & ABDMG (99.0\%)	        & 43.5~$\pm$~1.2$^{p}$  & 30, 4     \\
PSO J057.2893+15.2433       & PSO~057       & 03:49:09.44   & +15:14:36.00  & L7    & non-YMGM               & 25.2~$\pm$~3.0        & 31        \\
2MASS J04473039-1216155     & PSO~071 	    & 04:47:30.40   & -12:16:15.54  & T2    & non-YMGM               & 43.5~$\pm$~8.9$^{p}$  & 31, 2     \\
2MASS J21140802-2251358     & PSO~318       & 21:14:08.03   & -22:51:35.84  & L7    & BPMG (99.0\%)          & 22.2~$\pm$~0.9$^{p}$  & 32, 28    \\
2MASS J21171431-2940034     & PSO~319       & 21:17:14.31   & -29:40:03.42  & T0    & non-YMGM               & 19.1~$\pm$~2.8$^{p}$  & 31, 2     \\
2MASS J11101001+0116130     & SDSS~1110     & 11:10:10.01   & +01:16:13.09  & T5    & ABDMG (87.4\%)	        & 19.2~$\pm$~0.5$^{p}$  & 33, 34    \\
SIMP J013656.5+093347.3     & SIMP~0136     & 01:36:56.56   & +09:33:47.31  & T2    & CARN (96.3\%)	        & 6.1~$\pm$~0.1$^{p}$   & 35, 4     \\
ULAS J004757.40+154641.4    & ULAS~0047     & 00:47:57.32   & +15:46:41.56  & T2    & non-YMGM   	        & 37.0~$\pm$~4.5$^{p}$  & 36, 37    \\
ULAS J131610.13+031205.5    & ULAS~1316     & 13:16:10.13   & +03:12:05.56  & T3    & CARN (95.9\%)	        & 34.5~$\pm$~3.1$^{p}$  & 38, 2     \\
SIPS J1256-1257B            & VHS~1256~b    & 12:56:01.83   & -12:57:27.69  & L7    & non-YMGM   	        & 22.2~$\pm$~1.3$^{p}$  & 39, 40    \\
WISE J024124.73-365328.0    & WISE~0241     & 02:41:24.74   & -36:53:28.09  & T7    & ARG (91.5\%)	        & 19.1~$\pm$~1.0$^{p}$  & 41, 5     \\
WISE J052857.68+090104.4    & WISE~0528     & 05:28:57.69   & +09:01:04.40  & L1    & non-YMGM               & 81~$\pm$~13           & 42        \\
2MASS J08195820-0335266     & WISE~0819     & 08:19:58.21   & -03:35:26.65  & T4    & BPMG (83.9\%)	        & 13.9~$\pm$~0.6$^{p}$  & 43, 2     \\
PSO J334.1193+19.8800       & WISE~2216     & 22:16:28.62   & +19:52:48.10  & T3    & non-YMGM               & 30.0~$\pm$~11.2       & 31        \\
\hline
\hline
\end{tabular}
\tablefoot{$^{p}$: Distance calculated from parallax measurements.\\
\textbf{References.} 1. \cite{Kirkpatrick00}; 
2. \cite{Best20}
3. \cite{Wilson03}
4. \cite{Gaia20}
5. \cite{Kirkpatrick21}
6. \cite{Gagne15a}
7. \cite{Artigau15}
8. \cite{Dupuy18}
9. \cite{Dahn02}
10. \cite{Faherty09}
11. \cite{Cruz07}
12. \cite{Reid06}
13. \cite{Gagne18}
14. \cite{Cruz03}
15. \cite{Best21}
16. \cite{Cushing09}
17. \cite{Faherty12}
18. \cite{Schneider17}
19. \cite{Gagne17}
20. \cite{Kellogg15}
21. \cite{Best17}
22. \cite{Faherty16}
23. \cite{Gagne14}
24. \cite{Kendall04}
25. \cite{Knapp04}
26. \cite{Burgasser06}
27. \cite{Smart13}
28. \cite{Liu16}
29. \cite{Reid08}
30. \cite{Aller16}
31. \cite{Best15}
32. \cite{Liu13}
33. \cite{Geballe02}
34. \cite{Tinney14}
35. \cite{Artigau06}
36. \cite{Day13}
37. \cite{ZZhang21}
38. \cite{Marocco15}
39. \cite{Gauza15}
40. \cite{Dupuy20}
41. \cite{Tinney18}
42. \cite{Burgasser16}
43. \cite{Kirkpatrick11}

}
\end{table*}

\section{Semi-empirical analysis of the physical properties}
\label{sec:semiempirical_analysis}

By definition, because they have sub-stellar masses, brown dwarfs cannot trigger hydrogen fusion in their cores. This lack of internal energy production leads to cooling over time due to radiative losses. Evolutionary models, such as \texttt{COND03} \citep{Baraffe03}, predict how the fundamental properties of brown dwarfs evolve as a result of this cooling. For instance, the object contracts over time, leading to a decrease in radius. Since mass is assumed to remain constant with age, this contraction results in an increase in surface gravity. As a direct consequence of the decrease in \Teff and radius, the bolometric luminosity (\Lbol) also decreases with age, following the Stefan-Boltzmann law.

In this section, we will estimate the age and \Lbol of the X-SHYNE objects and inject these values into the \texttt{COND03} evolutionary model to derive other fundamental parameters. We will first discuss the spectral types of our sample that are also strongly related to the \Teff.

\subsection{A wide range of spectral types covered}
More than two decades ago, the empirical exploration of the morphology of brown dwarf spectra has led to the definition of the L \citep{Kirkpatrick99, Martin99} and T \citep{Burgasser02T, Geballe02} sequences. It was observed that with decreasing temperature, certain absorptions evolved, as did the overall shape of the pseudo-continuum (see Figures \ref{fig:sequence_full_wav}, \ref{fig:sequence_VIS_wav}-\ref{fig:sequence_K_wav}). These classifications have allowed the identification of template spectra libraries used to determine the spectral type of newly discovered objects. However, this method has its limitations. \cite{Cruz18} demonstrated the differences that can exist between the spectral type determined from atomic (Cs, Rb) and molecular (VO, TiO, CrH, FeH) absorptions detected in the visible range ($\sim$0.6-0.9~\mic) and the spectral type determined from near-infrared data ($\sim$1.0-2.5~\mic), which are dominated by the effects of gravity, metallicity, and clouds condensation. Due to biases in the spectral type determination method and to remain conservative, we decided to assign an uncertainty of one spectral subclass to the spectral types of the X-SHYNE objects listed in Table \ref{Tab:sample_ymg}. As a result, X-SHYNE covers a range of spectral types from L0$\pm$1 to T7$\pm$1.

\begin{figure}[h!]
\centering
\includegraphics[width=1.\hsize]{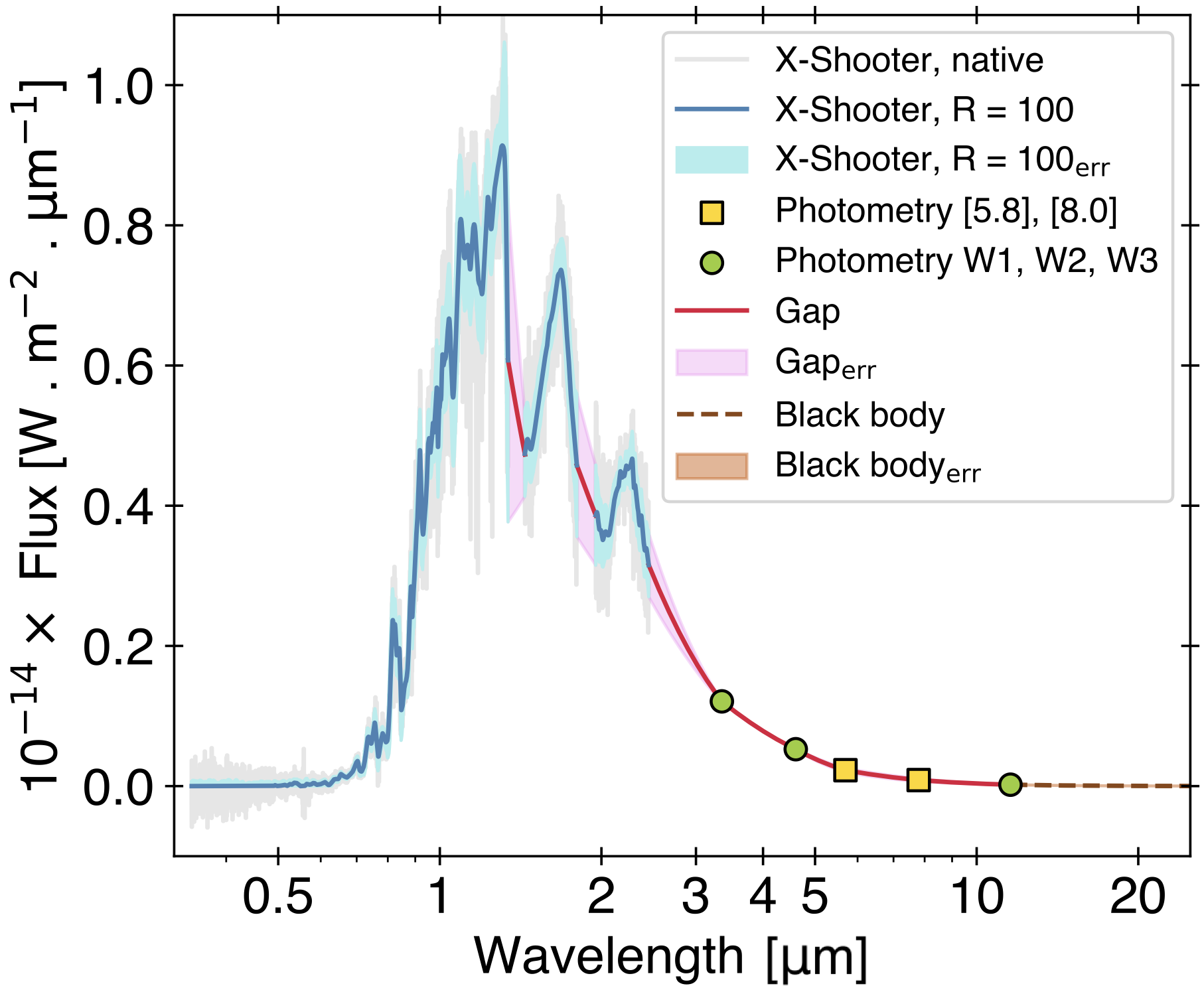}
    \caption{Reconstruction of the Spectral Energy Distribution (SED). We have adapted the procedure described in \cite{Filippazzo15} to combine our X-Shooter spectra with previously obtained photometric data. This reconstructed SED was used to calculate the bolometric luminosity of our targets. The example presented here is 2MASS~0046.}
    \label{fig:ex_sed_reconstructed}
\end{figure}

\subsection{Age determination}
\label{sec:semiempirical_age}

Stellar associations are ideal environments for the search and the characterization of low-mass brown dwarfs as the objects are comparatively brighter than those found in star-forming regions \citep{Lucas01, Lodieu18, McCaughrean23}. Objects within these groups are assumed to share similar kinematic properties and age, under the hypothesis that they formed simultaneously in the same region. Age determination, in particular, currently relies on several diagnostics (see \citealt{Gagne24} for a recent review). These include predictions from stellar and sub-stellar internal structure models, often represented in color-magnitude diagrams (photometric isochrones, \citealt{Nordstrom04}); measurements of stellar and sub-stellar rotation for a given spectral type (gyro-chronology, \citealt{Barnes09}; \citealt{Mamajek09}); spectroscopic indicators of chromospheric activity or accretion, such as the H$_{\alpha}$ emission line; the presence of elements that are destroyed over time (e.g., lithium, \citealt{Michaud91}; \citealt{Rebolo91}); X-ray activity; and the detection of infrared excess \citep{Skumanich72, Soderblom91}.

\begin{table}[t]
\centering
\caption{Properties of the young associations considered here.}
\label{Tab:mov_group}
\small
\begin{tabular}{l|ll|c|l}
\hline
Name        &  Distance             & Age                   &  N$_{X-SHYNE}$    & Ref.      \\
            & [pc]                  &  [Myr]                &                   &           \\ 
\hline
ABDMG       &  $\sim$20             &  149$_{-51}^{+19}$    & 8           & 1, 2          \\ 
ARG         &  $\sim$140            &  $\sim$40             & 2           & 2, 3          \\ 
BPMG        &  $\sim$35             &  24$\pm$3             & 4           & 2, 4          \\ 
CARN        &  $\sim$30             &  $\sim$200            & 4           & 5             \\ 
COL         &  $\sim$70             &  42$_{-4}^{+6}$       & 1           & 2, 6          \\ 
CRIUS203    & -                     & $\sim$500             & 2           & 6, Priv. com.\\
Tuc-Hor     &  $\sim$50             &  45$\pm$4             & 4           & 2, 6         \\ 
TWA         &  $\sim$65             &  10$\pm$3             & 1           & 2, 6          \\ 
\hline 
non-YMGM       &  -                    &  -                 & 17          & -         \\ 
\hline
\hline
\end{tabular}
\tablefoot{The number of X-SHYNE objects associated with each group is given. 17 sources are not identified as members of a young moving group (non-YMGM). Full names: AB Doradus : ABDMG; Argus : ARG; $\beta$ Pictoris : BPMG; Carina-Near : CARN; Columba : COL; Tucana Horologium : Tuc-Hor; TW Hydrae : TWA.\\ 
\textbf{References.} 1. \cite{Zuckerman04}; 2. \cite{Bell15}; 3. \cite{Torres08}; 4. \cite{Zuckerman01}; 5. \cite{Zuckerman06}; 6. \cite{Moranta22}.}
\end{table}

\begin{table*}[t]
\centering
\caption{Properties of the X-SHYNE library from the semi-empirical analysis.}
\label{Tab:semi_emp_param}
\small
\begin{tabular}{l|l|llll}
\hline
Target     &  log(L/L$_{\odot}$)           & Mass              & \Teff         & Radius        &  log(g)                   \\
           &                            & [\Mjup]           &  [K]          & [\Rjup]       &  [dex]                      \\ 
\hline
2MASS 0030 & -4.44~$\pm$~0.21 &  - &  - &  - &  - \\ 
2MASS 0045 & -3.63~$\pm$~0.06 & 19.1$_{-4.96}^{+0.96}$ & 1885$_{-90}^{+18}$ & 1.49$_{-0.11}^{+0.08}$ & 4.35$_{-0.15}^{+0.08}$ \\ 
2MASS 0046 & -3.26~$\pm$~0.06 &  - &  - &  - &  - \\ 
2MASS 0103 & -4.2~$\pm$~0.24 & 44.01$_{-10.22}^{+10.43}$ & 1641$_{-221}^{+247}$ & 1.0$_{-0.0}^{+0.05}$ & 5.06$_{-0.15}^{+0.09}$ \\ 
2MASS 0153 & -3.89~$\pm$~0.11 & 14.19$_{-2.0}^{+3.53}$ & 1647$_{-95}^{+122}$ & 1.42$_{-0.05}^{+0.06}$ & 4.25$_{-0.05}^{+0.11}$ \\ 
2MASS 0219 b & -3.84~$\pm$~0.21 & 15.09$_{-3.13}^{+5.59}$ & 1691$_{-173}^{+217}$ & 1.43$_{-0.07}^{+0.11}$ & 4.27$_{-0.07}^{+0.16}$ \\ 
2MASS 0249 c & -3.99~$\pm$~0.28 &  - &  - &  - &  - \\ 
2MASS 0326 & -3.8~$\pm$~0.15 & 32.53$_{-8.06}^{+8.81}$ & 1887$_{-173}^{+169}$ & 1.2$_{-0.03}^{+0.06}$ & 4.77$_{-0.15}^{+0.1}$ \\ 
2MASS 0342 & -3.51~$\pm$~0.11 & 15.83$_{-1.15}^{+8.99}$ & 1941$_{-92}^{+168}$ & 1.57$_{-0.16}^{+0.07}$ & 4.22$_{-0.02}^{+0.21}$ \\ 
2MASS 0355 & -4.13~$\pm$~0.05 & 24.51$_{-6.62}^{+3.93}$ & 1582$_{-98}^{+72}$ & 1.18$_{-0.02}^{+0.05}$ & 4.65$_{-0.17}^{+0.08}$ \\ 
2MASS 0508 & -3.68~$\pm$~0.07 & 36.68$_{-7.58}^{+6.35}$ & 2006$_{-109}^{+93}$ & 1.22$_{-0.03}^{+0.05}$ & 4.8$_{-0.12}^{+0.08}$ \\ 
2MASS 0512 & -4.15~$\pm$~0.11 &  - &  - &  - &  - \\ 
2MASS 0518 & -3.42~$\pm$~0.09 & 22.19$_{-7.13}^{+5.68}$ & 2051$_{-120}^{+134}$ & 1.62$_{-0.08}^{+0.13}$ & 4.36$_{-0.18}^{+0.12}$ \\ 
2MASS 0616 & -3.52~$\pm$~0.64 &  - &  - &  - &  - \\ 
2MASS 0723 & -3.92~$\pm$~0.13 & 32.41$_{-4.24}^{+5.72}$ & 1798$_{-121}^{+132}$ & 1.17$_{-0.02}^{+0.02}$ & 4.79$_{-0.07}^{+0.06}$ \\ 
2MASS 1021 & -4.14~$\pm$~0.24 &  - &  - &  - &  - \\ 
2MASS 1119 & -4.41~$\pm$~0.36 &  - &  - &  - &  - \\ 
2MASS 1147 & -4.21~$\pm$~0.14 &  - &  - &  - &  - \\ 
2MASS 1148 & -3.89~$\pm$~0.19 &  - &  - &  - &  - \\ 
2MASS 1207 & -3.39~$\pm$~0.13 & 13.38$_{-3.12}^{+1.4}$ & 1985$_{-131}^{+98}$ & 1.71$_{-0.08}^{+0.13}$ & 4.07$_{-0.11}^{+0.02}$ \\ 
2MASS 1213 & -4.17~$\pm$~0.07 & 26.31$_{-2.75}^{+2.72}$ & 1573$_{-75}^{+71}$ & 1.16$_{-0.01}^{+0.01}$ & 4.7$_{-0.05}^{+0.05}$ \\ 
2MASS 1425 & -4.04~$\pm$~0.05 & 26.67$_{-5.71}^{+3.9}$ & 1667$_{-92}^{+69}$ & 1.19$_{-0.02}^{+0.04}$ & 4.69$_{-0.12}^{+0.08}$ \\ 
2MASS 1521 & -4.75~$\pm$~0.31 &  - &  - &  - &  - \\ 
2MASS 1826 & -3.76~$\pm$~0.21 & 12.11$_{-1.81}^{+1.81}$ & 1723$_{-162}^{+160}$ & 1.53$_{-0.06}^{+0.09}$ & 4.13$_{-0.04}^{+0.01}$ \\ 
2MASS 2104 & -3.7~$\pm$~0.15 & 12.5$_{-1.32}^{+1.41}$ & 1767$_{-114}^{+114}$ & 1.54$_{-0.05}^{+0.08}$ & 4.14$_{-0.03}^{+0.01}$ \\ 
2MASS 2139 & -4.76~$\pm$~0.1 & 28.82$_{-4.65}^{+3.21}$ & 1182$_{-69}^{+73}$ & 1.01$_{-0.01}^{+0.04}$ & 4.85$_{-0.11}^{+0.06}$ \\ 
2MASS 2206 & -3.95~$\pm$~0.09 & 28.69$_{-6.61}^{+5.82}$ & 1747$_{-128}^{+111}$ & 1.19$_{-0.03}^{+0.05}$ & 4.72$_{-0.14}^{+0.09}$ \\ 
2MASS 2244 & -4.51~$\pm$~0.08 & 12.61$_{-0.92}^{+7.96}$ & 1234$_{-51}^{+120}$ & 1.24$_{-0.09}^{+0.03}$ & 4.32$_{-0.04}^{+0.28}$ \\ 
2MASS 2322 & -3.73~$\pm$~0.09 & 17.17$_{-3.9}^{+3.37}$ & 1791$_{-103}^{+110}$ & 1.46$_{-0.07}^{+0.08}$ & 4.31$_{-0.11}^{+0.11}$ \\ 
2MASS 2354 & -3.74~$\pm$~0.09 & 34.7$_{-7.44}^{+6.88}$ & 1949$_{-125}^{+113}$ & 1.21$_{-0.03}^{+0.05}$ & 4.79$_{-0.13}^{+0.09}$ \\ 
PSO 057 & -4.58~$\pm$~0.18 &  - &  - &  - &  - \\ 
PSO 071 & -4.29~$\pm$~0.31 &  - &  - &  - &  - \\ 
PSO 318 & -4.52~$\pm$~0.09 & 6.58$_{-0.91}^{+1.02}$ & 1177$_{-59}^{+61}$ & 1.39$_{-0.02}^{+0.02}$ & 3.94$_{-0.07}^{+0.06}$ \\ 
PSO 319 & -4.46~$\pm$~0.22 &  - &  - &  - &  - \\ 
SDSS 1110 & -5.01~$\pm$~0.18 & 11.63$_{-2.94}^{+0.32}$ & 961$_{-116}^{+105}$ & 1.18$_{-0.01}^{+0.03}$ & 4.33$_{-0.14}^{+-0.0}$ \\ 
SIMP 0136 & -4.73~$\pm$~0.27 & 12.28$_{-1.1}^{+8.71}$ & 1119$_{-152}^{+222}$ & 1.21$_{-0.07}^{+-0.02}$ & 4.33$_{-0.02}^{+0.29}$ \\ 
ULAS 0047 & -4.71~$\pm$~0.31 &  - &  - &  - &  - \\ 
ULAS 1316 & -4.46~$\pm$~0.25 & 19.86$_{-7.7}^{+6.82}$ & 1331$_{-205}^{+224}$ & 1.16$_{-0.0}^{+0.06}$ & 4.58$_{-0.26}^{+0.13}$ \\ 
VHS 1256 b & -4.46~$\pm$~0.55 & 13.37$_{-2.38}^{+16.72}$ & 1265$_{-320}^{+518}$ & 1.24$_{-0.05}^{+-0.03}$ & 4.35$_{-0.04}^{+0.39}$ \\ 
WISE 0241 & -5.3~$\pm$~0.53 & 4.82$_{-2.51}^{+4.07}$ & 776$_{-204}^{+270}$ & 1.27$_{-0.03}^{+0.04}$ & 3.88$_{-0.32}^{+0.28}$ \\ 
WISE 0528 & -3.55~$\pm$~0.22 &  - &  - &  - &  - \\ 
WISE 0819 & -4.79~$\pm$~0.31 & 5.04$_{-1.68}^{+2.26}$ & 1019$_{-163}^{+191}$ & 1.37$_{-0.03}^{+0.04}$ & 3.84$_{-0.16}^{+0.15}$ \\ 
WISE 2216 & -4.75~$\pm$~0.54 &  - &  - &  - &  - \\ 
\hline
\hline
\end{tabular}
\end{table*}

Most of the X-SHYNE objects were selected as Young co-Moving Group Members (YMGM) or member candidates before the important Gaia DR2 release that significantly modified key kinematics properties for these objects. The properties of the groups considered in this work are summarized in Table \ref{Tab:mov_group}. Using updated astrometry and radial velocity measurements, we recalculated the membership probability for each X-SHYNE target with the \texttt{BANYAN $\Sigma$} tool \citep{Gagne18b}. These results, along with their associated probabilities, are summarized in Table \ref{Tab:sample_ymg}. For objects with a membership probability below 80\%, we classified them as Non-Young co-Moving Group Members (non-YMGM). Consequently, their age remains uncertain and requires further confirmation. In the specific case of the companion VHS~1256~b, although it is classified as a non-YMGM, the age of its primary star has been derived independently by \cite{Dupuy23} as 140$\pm$20~Myr and will be assumed as the age of the companion.

The equivalent width (EW) of the potassium lines at 1.169, 1.177, 1.244, and 1.253~\mic are known to be robust discriminants between old and young L- and M-type objects \citep{Cushing05, Bayo11, Allers13, Bonnefoy14, Martin17, Manjavacas20, Piscarreta24}. This is because their strength correlates with the pressure in the photosphere, which, in turn, is linked to the surface gravity of the object. The K~I EW of some X-SHYNE objects has previously been calculated and analyzed in \cite{Piscarreta24}. To ensure the homogeneity of our study, we recalculated these values using our framework.

To achieve this, we estimated the pseudo-continuum on both sides of the spectral line by calculating the median flux within the spectral ranges d$\rm \lambda_{cont. left}$ and d$\rm \lambda_{cont. right}$, as listed in Table \ref{Tab:ew_wav}. We then linearly interpolated these two flux values across the spectral range d$\rm \lambda_{line}$ defined for the line. Finally, we computed the equivalent width using the following equation:

\begin{equation}
  EW~=~\sum_{\lambda} \left(1-\frac{F_{line}}{F_{cont. interp}}\right) \times d\lambda
\end{equation}

where $F_{line}$ and $F_{cont. interp}$ represent the flux values of the spectral line and the interpolated pseudo-continuum, respectively, at the location of the line. The term $d\lambda$ corresponds to the wavelength grid of the data.

\begin{table}[h]
\centering
\caption{The wavelength ranges used to calculate the equivalent widths of each K~I line.}
\label{Tab:ew_wav}
\small
\begin{tabular}{c|c|c|c}
\hline
K~I lines     &  d$\rm \lambda_{cont. left}$   & d$\rm \lambda_{line}$  & d$\rm \lambda_{cont. right}$      \\
 (\mic)       &  (\mic)                        & (\mic)                 & (\mic)                            \\ 
\hline
1.169         & 1.155 - 1.167                  & 1.168 - 1.1705         & 1.1705, 1.175                     \\ 
1.177         & 1.1708 - 1.1757                & 1.1757 - 1.179         & 1.180 - 1.185                     \\ 
1.244         & 1.235 - 1.2417                 & 1.2417 - 1.2445        & 1.246 - 1.250                     \\ 
1.253         & 1.246 - 1.250                  & 1.250 - 1.255          & 1.255 - 1.265                     \\ 
\hline
\hline
\end{tabular}
\end{table}

We then compared the EW of the four K~I absorption lines in the X-SHYNE sources with those derived from the spectral library (\Rlam$\sim$2000) published by \cite{Martin17}. This library includes isolated brown dwarfs with low, medium, and high surface gravity, corresponding to young ($<$30~Myr), intermediate (30–100~Myr), and field ($>$100~Myr) ages, respectively. In their study, \cite{Martin17} calculated the EW following the method described by \cite{Allers13}, which determines the pseudo-continuum at each line location using a linear interpolation between two spectral windows located on either side of the line. To ensure a consistent comparison, we adopted the same technique. To estimate the error bars, we followed the procedure described in \cite{Petrus23}. Specifically, for each target, we constructed 100 spectra by randomly generating the flux at each wavelength using a Gaussian distribution, with the mean set to the initial flux value and the standard deviation equal to the flux error. The final EW value and its associated error were obtained as the mean and standard deviation of the 100 EWs, respectively.

Figure \ref{fig:ki_age} shows the evolution of the four EWs as a function of spectral type. We confirmed that the EWs measured for young objects are significantly lower than those of field objects for L-type dwarfs. This trend persists for T-dwarfs with the K~I lines at 1.169 and 1.177~\mic, but is less clear for the K~I lines at 1.243 and 1.253~\mic, which are blended with a FeH feature that can affect the EW measurement. \cite{Manjavacas20} and \cite{Martin17} established empirical relations between these EWs and the ages of objects at the M/L transition. These relations do not enable a reliable age estimation for the non-YMGM objects in X-SHYNE, which spans the L/T transition. Furthermore, our sample size is too limited to derive such relationships independently in that work. However, we confirm the young age of all X-SHYNE objects (10-500~Myr).

\begin{figure*}[h!]
\centering
\includegraphics[width=1.0\hsize]{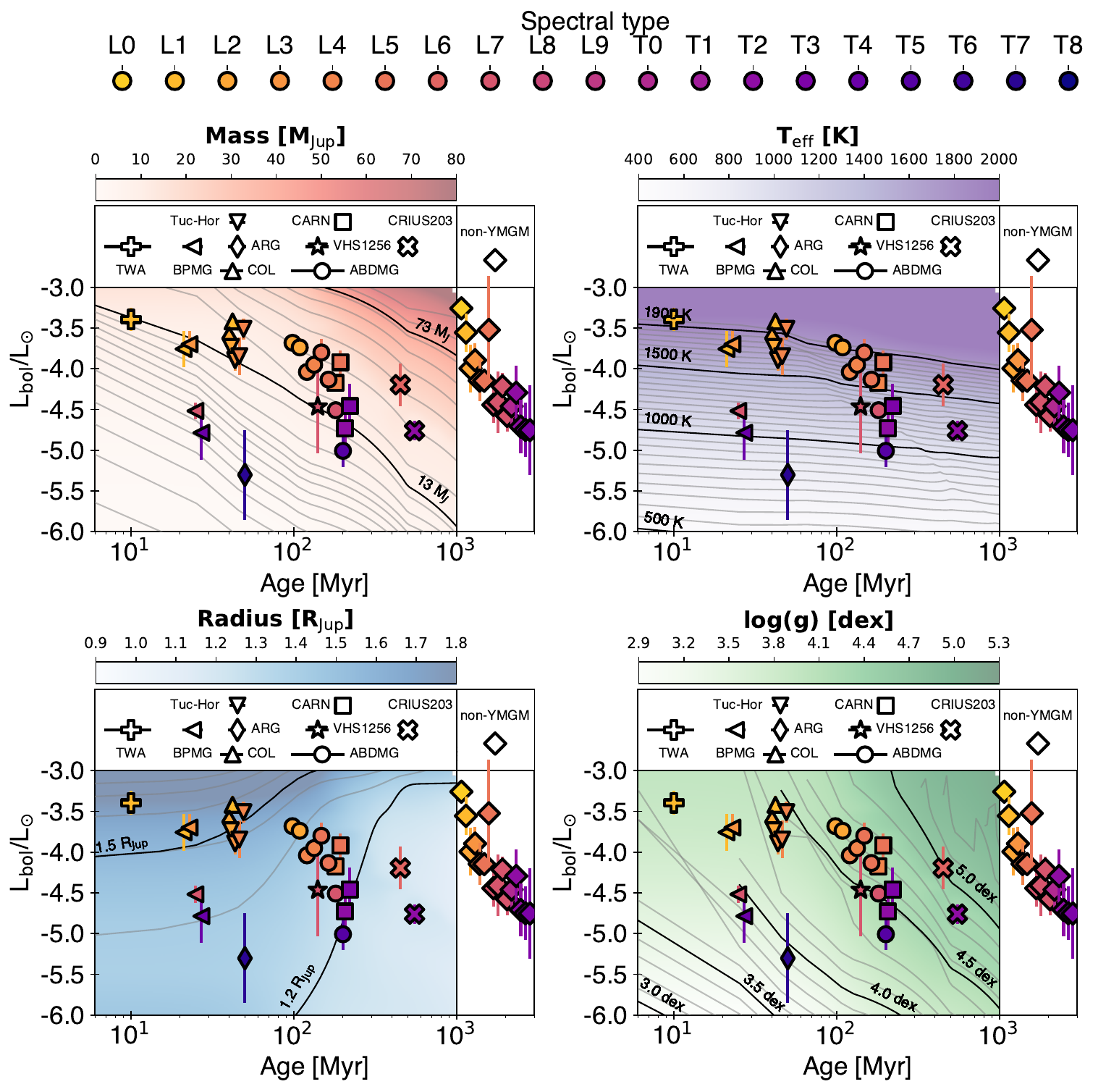}
    \caption{Prediction of the evolutionary model \texttt{COND03} \citep{Baraffe03}. We injected the age estimated in Section \ref{sec:semiempirical_age} and the \Lbol calculated in Section \ref{sec:semiempirical_Lbol} into the model, which we interpolated linearly to estimate the mass (top-left panel, red background), \Teff (top-right panel, purple background), radius (bottom-left panel, blue background), and log(g) (bottom-right panel, green background). The non-YMGM objects, for which we do not have an age estimate, are plotted on the right side of each panel. For each parameter, we also display its variation as a function of \Lbol and age (color bar and solid gray and black lines). The shape of each X-SHYNE dot represents its membership, while its color corresponds to its spectral type. To improve the clarity of the plot, objects that share the same age have been evenly distributed within a range corresponding to the error bars of that age.}
    \label{fig:recap_lum_age}
\end{figure*}

\subsection{Bolometric luminosity determination}%\Lbol}
\label{sec:semiempirical_Lbol}

The calculation of \Lbol is not straightforward because it requires integrating the emitted flux of an object over its entire spectral energy distribution (SED), while available data often only cover a fraction of this SED. Atmospheric models are typically used to fill these wavelength gaps (e.g., \citealt{Stephens09}; \citealt{Dieterich14}; \citealt{Carter23}), but their systematic errors, along with the difficulty of propagating these uncertainties into the error bars of \Lbol, may bias its calculation and interpretation. Bolometric corrections, which establish a relationship between \Lbol and specific photometric points, can also be employed (e.g., \citealt{Golimowski04}; \citealt{Vrba04}; \citealt{Zapatero14}). However, like atmospheric models, this method suffers from systematic errors and significant dispersion, particularly for late-L and T-type objects. To provide a more robust estimate of \Lbol, \cite{Filippazzo15} proposed a method that reconstructs the full SED by combining available spectroscopic and photometric data. For some X-SHYNE targets, this approach has already been applied by \cite{Filippazzo15}, \cite{Faherty16} and more recently by \cite{2023ApJ...959...63S}. By adding new spectra obtained from X-Shooter, we have decided to recalculate \Lbol for every targets, homogeneously, following the same methodology. Figure \ref{fig:ex_sed_reconstructed} illustrates the various contributions to the \Lbol calculation, using 2MASS~0046 (L0) as a case study.

\subsubsection{Spectroscopic data}
We used X-Shooter data covering the wavelength range between 0.345 and 2.45~\mic. To calculate a robust value of \Lbol and mitigate the noise that can affect high-resolution data, we focused on the pseudo-continuum by reducing the spectral resolution to \Rlam$\sim$100 by applying a Gaussian filter. The new error bars were estimated as the standard deviation of the native spectrum within a box of 1000 spectral channels centered around each point. To avoid contamination from the broad water absorption bands, we selected only three spectral regions: SHORT (0.345–1.340~\mic), H (1.44–1.80~\mic), and K (1.952–2.450~\mic). Due to data quality issues, likely caused by slit misalignment during observations, the SHORT-band was restricted to 1.01–1.34~\mic\ for 2MASS~1826 and to 0.6–1.34\mic\ for ULAS~1316. Similarly, the K-band for 2MASS~0508 was limited to 1.952–2.22~\mic.

\subsubsection{Photometric data}
We considered five different photometric points. The W1 ($\rm \Delta\lambda$ = 2.754-3.872~\mic), W2 ($\rm \Delta\lambda$ = 3.963-5.341~\mic), and W3 ($\rm \Delta\lambda$ = 7.443-17.261~\mic) values and their associated errors, were retrieved from \cite{Marocco21} and \cite{Cutri12}. For 2MASS~0219~b, SIMP~0136, VHS~1256~b, WISE~0241, and WISE~0819, where WISE photometry was unavailable, we estimated the photometric values using the empirical relation between spectral type and WISE photometry defined by \cite{Faherty16}. The uncertainties were propagated by combining the error from the relation itself with an additional uncertainty of 1 spectral subtype.

In addition to the WISE data points, the IRAC [5.8~\mic] and [8~\mic] photometry were considered for reconstructing the SED. However, since these IRAC measurements were not available for most of our targets, we chose to estimate them using semi-empirical relations between these two IRAC bands and WISE photometry, as established by \cite{Filippazzo15}. This approach ensures a more robust estimate of the SED at these wavelengths compared to simple interpolation, as described in Section \ref{Sec:gaps}. These relations differ slightly between young and old objects and are only valid within a specific range of W1 magnitudes. We prioritized the relation defined for young objects; however, for WISE~0241 and WISE~0819, we used the relation for old objects because their W1 photometry fell outside the valid range for young-object relations. Nevertheless, since the flux values at these wavelengths are very low, this choice has only a minimal impact on the \Lbol calculation.  As before, the uncertainties were propagated by combining the error from the relation itself with the uncertainty associated with the considered W1 magnitude.

\subsubsection{Gaps definition} 
\label{Sec:gaps}
To construct a complete SED covering the entire wavelength range, we filled the gaps between the different datasets. The water bands, as well as the gaps between photometric points, were filled using a linear interpolation of the logarithms of the fluxes on either side of the gaps. This approach mimics the behavior of a blackbody law at these wavelengths. We also fitted a blackbody law to the photometric points and used it to extend the SED between the W3 photometric point and 1 mm. At wavelengths shorter than 0.360~\mic, we assumed zero flux, since this spectral range—uncovered by X-Shooter—contributes less than 0.01\% of the total flux, according to models of atmosphere. For 2MASS~1826, an additional blackbody law was fitted using the wavelength range between 1.000 and 1.345\mic to fill the SED between 0.36 and 1.00~\mic. The flux gaps were computed for the mean data flux, as well as for the minimum and maximum data flux values, by propagating the applying uncertainties. This allowed us to define the errors at the locations of the gaps.

To calculate the \Lbol we integrated the spectroscopic data with the reduced resolution and the gaps between 0.346 and 1000~\mic considering negative flux as null.

\subsection{Masses, temperatures and radii predicted by evolutionary models}
\label{sec:semiempirical_modelEvol}
For the YMGMs of X-SHYNE, we used our age estimates (defined in Section~\ref{sec:semiempirical_age}) and our \Lbol estimates (defined in Section~\ref{sec:semiempirical_Lbol}) as inputs into the \texttt{COND03} evolutionary model \citep{Baraffe03}. We linearly interpolated this model to estimate the radius, mass, \Teff, and log~g of the objects. The behavior of these parameters within the \texttt{COND03} model, as well as the predictions for the X-SHYNE sample, are illustrated in Figure \ref{fig:recap_lum_age}. This semi-empirical analysis confirms the impact of the \Teff on the spectral type as identified in many other studies (e.g. \citealt{Kirkpatrick99}, \citealt{Burgasser02T}).
It also confirms that a young early-L dwarf can have the same mass as an older T-type dwarf. Additionally, we notice that the predicted radii of old, low-mass objects, which are expected to reach values below 1~\Rjup, show significant dispersion. The luminosity, mass, \Teff, radius, and log(g) values for the X-SHYNE YMGMs estimated from this semi-empirical analysis are summarized in Table \ref{Tab:semi_emp_param}. These results will be compared to those derived in the next section, which presents our synthetic analysis.

\begin{figure*}[h!]
\centering
\includegraphics[width=1.0\hsize]{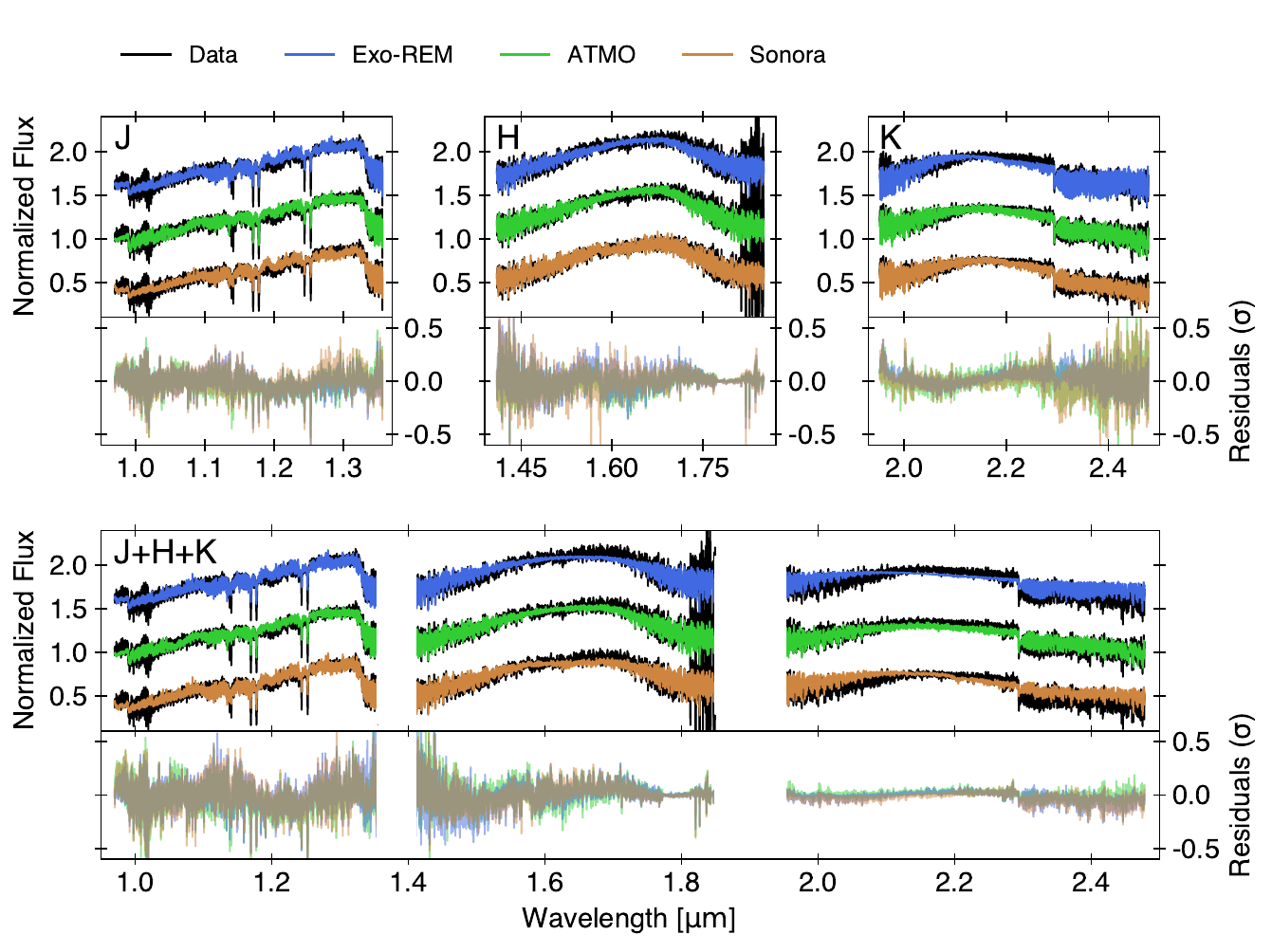}
    \caption{Comparison between the X-Shooter spectra and the best fit obtained with each model when considering the J band (top-left panel), H band (top-center panel), and K band (top-right panel), as well as the combination of all bands (bottom panel). The example shown here is 2MASS~0046.}
    \label{fig:example_best_fit}
\end{figure*}

\begin{table*}[t]
\centering
\caption{Priors used for the fits with the three synthetic spectra grids considered}
\label{Tab:priors}
\small
\begin{tabular}{ll|lll}
\hline
Param.            &       &  \texttt{Exo-REM}           & \texttt{Sonora}               & \texttt{ATMO}                 \\
\hline
\Teff             & [K]   &  $\mathcal{U}(400, 2000)$    &  $\mathcal{U}(900, 2400)$    &  $\mathcal{U}(800, 3000)$     \\ 
log(g)            & [dex] &  $\mathcal{U}(3.0, 5.0)$     &  $\mathcal{U}(3.5, 5.5)$     &  $\mathcal{U}(2.5, 5.5)$      \\ 
$\rm [M/H]$       &       &  $\mathcal{U}(-0.5, 0.5)$    &  $\mathcal{U}(-0.5, 0.5)$    &  $\mathcal{U}(-0.6, 0.6)$     \\ 
C/O               &       &  $\mathcal{U}(0.1, 0.8)$     &  -                           &  $\mathcal{U}(0.3, 0.7)$      \\ 
\fsed             &       &  -                           &  $\mathcal{U}(1.0, 8.0)$     &  -        \\ 
$\rm \gamma$      &       &  -                           &  -                           &  $\mathcal{U}(1.01, 1.05)$    \\ 
\hline 
RV                        & [km.s$^{-1}$] &  $\mathcal{U}(-100, 100)$    &  $\mathcal{U}(-100, 100)$    &  $\mathcal{U}(-100, 100)$     \\ 
R                         & [\Rjup] &  $\mathcal{U}(0, 10)$        &  $\mathcal{U}(0, 10)$        &  $\mathcal{U}(0, 10)$         \\ 
d                         & [pc] &  $\mathcal{N}(\rm d_{i},~d^{2}_{err, i})$  &  $\mathcal{N}(\rm d_{i},~d^{2}_{err, i})$  &  $\mathcal{N}(\rm d_{i},~d^{2}_{err, i})$     \\ 
\hline
\hline
\end{tabular}
\tablefoot{$\rm d_{i}$ and $\rm d^{2}_{err, i}$ are the distance and its error, respectively, given in Table \ref{Tab:sample_ymg}.}
\end{table*}

\section{Synthetic atmosphere analysis}
\label{sec:syntetic_analysis}

Recent work led by \cite{Petrus20, Petrus23, Petrus24} and \cite{Palma23} have confirmed the results from \cite{Marocco14} who shown the limitations of fitting models using wide wavelength coverages, demonstrating that atmospheric models fail to reproduce the data globally. Considering a diversity of models, they have highlighted the interest of independent fits by selecting restricted wavelength windows that can be accurately reproduced by the models. In particular, \cite{Petrus24} proposed a method to propagate part of the systematic errors of the models into the final parameter errors by exploiting the dispersion of these estimations for different spectral windows considered for the independent fits. However, this technique has its own limitations. It is still dependent of the quality of the models and their ability to reproduce correctly the spectral feature. For instance \cite{Petrus24} noticed that in the case of VHS~1256~b, nonphysical parameters estimated at the location of a silicate absorption ($\sim$10~\mic) which was known to be not reproduced by the models, could bias the final results. Therefore, the results are dependent of the choice of the spectral windows considered. In this work, we propose another approach to take into account the systematic errors of the models.

\subsection{The atmospheric models}
\label{subsec:formosa_models}
In this study, we considered three different grids of synthetic spectra. The \texttt{Exo-REM} model \citep{Charnay18} incorporates non-equilibrium chemistry between CO, CH$_{4}$, CO$_{2}$, and NH$_{3}$, and the formation of clouds composed of iron, Na$_{2}$S, KCl, silicates, and water. The \texttt{Sonora-diamondback} model (hereafter \texttt{Sonora}) \citep{Morley24} assumes chemical equilibrium throughout the atmosphere and includes a cloud model parameterized by a sedimentation factor \fsed. The \texttt{ATMO} model \citep{Tremblin15} simulates the impact of clouds on the spectrum through diabatic convection (fingering convection) led by the adiabatic index $\rm \gamma$ generated by non-equilibrium chemical reactions: CO/CH$_{4}$ at the L-T transition and N$_{2}$/NH$_{3}$ at the T-Y transition. A detailed description of these three models can be found in \cite{Petrus24}.

\subsection{Optimized fitting strategy}
\label{subsec:formosa_meth}

The fits are performed using the code \texttt{ForMoSA}\footnote{https://formosa.readthedocs.io/}, which is based on a Bayesian algorithm known as nested sampling. This algorithm allows for a comprehensive exploration of the parameter space defined by the grids of synthetic spectra. This exploration is facilitated by calculating a likelihood function at each iteration, which is used to define N-dimensional iso-surfaces (where N is the number of parameters being explored) that converge towards the maximum likelihood. In this analysis, we assume Gaussian and independent errors for the data, so maximizing the likelihood function is equivalent to minimizing a $\rm \chi ^{2}$ function. This assumption may be incorrect for spectra that should include correlated noise, but to the best of our knowledge, no pipeline fully tracks the covariance to the science-ready spectra. In addition to the parameters considered by the grids, we extend the parameter space by exploring the radial velocity, which is performed by applying a Doppler shift to the synthetic spectra. The radius and distance are also constrained, serving as a dilution factor $\rm C_{K} = (R/d)^{2}$ between the synthetic flux generated at radius R and the data flux observed at distance d. A normal law is used as prior on the distance, with the mean value and standard deviation provided in Table \ref{Tab:sample_ymg}. The other parameters are initialized with uniform priors summarized in Table \ref{Tab:priors}. We used the nested sampling Python package \texttt{pyMultinest} \citep{Buchner14} and injected 200 living points during the inversion.

To evaluate the performance of models across the spectral types covered by X-SHYNE, we adapted the methodology of \cite{Petrus24} to VLT/X-Shooter data, analyzing three distinct spectral windows independently. These windows correspond to the J (0.97–1.35~\mic), H (1.41–1.85~\mic), and K (1.95–2.48~\mic) infrared bands. Additionally, we performed a combined fit across all three bands simultaneously, denoted as J+H+K. Due to limitations in the ability of the models to reproduce wavelengths shorter than 1~\mic \citep{Petrus23}, this range was excluded from our analysis. Unlike the approach in \cite{Petrus24}, where the dispersion of the model predictions along the entire SED was exploited, we employed an alternative strategy. This assumes that the observed data error bars are insufficient to fully account for the discrepancies between the data and models. Our fitting routine consists of iterative loops that include the error inflation for the data, the spectral inversion, the computation of the reduced $\chi^2_{red}$ between the data and the best-fitting model, and the recalibration of the error inflation factor to scale $\chi^2_{red}$ to 1. If, after a spectral inversion, $\chi^2_{red}$ lies within [0.5, 1.5], the loop terminates, and the parameters are extracted from this final fit. Otherwise, a new iteration begins with the rescaled error bars. The spectral inversion employs a Bayesian algorithm, and the inflated data errors, which reflect the models' inability to fully reproduce the observations, are propagated into the parameter uncertainties. This concept aligns with the methodology implemented in the \texttt{Starfish} code \citep{Czekala15, Gully17}, optimized to analyze stellar spectra, which calculates a covariance matrix for model grids using Gaussian process kernels and incorporates it into the likelihood calculation (\citealt{Zzzhang21}, \citealt{Zzzhang21_2}). While \texttt{Starfish} applies this as wavelength-dependent error inflation for each spectral channel, we simplified the approach by using a constant inflation factor across the wavelength range. Incorporating the calculation of model covariance matrices into the \texttt{ForMoSA} framework will be addressed in a future update.

Despite optimizing our fitting strategy to ensure consistency between the synthetic spectra and the data, the posterior distribution of each parameter remains tightly constrained, resulting in unrealistically small uncertainties of $<$1~K for \Teff and $<$0.01 for log(g) and chemical abundances. Such small errors make the fit overly sensitive to minor irregularities in the model grid, potentially causing the posteriors to align with grid nodes and introducing biases.

To address this, we arbitrarily applied an inflation factor of 10, leading to revised uncertainties $>$10~K for \Teff and $>$0.05 for log(g) and chemical abundances. Since our analysis relies on precomputed grids of synthetic spectra, the fit results are primarily dictated by self-consistent model quality. Thus, inflating the observational errors does not affect the mean values of the posteriors but rather broadens their standard deviations. Furthermore, even with this inflation factor, our derived parameter uncertainties remain lower than those estimated in \cite{Petrus24}, where the parameter dispersion along the SED was considered. Consequently, the uncertainties on the parameters should be interpreted with caution, as they are likely still dominated by model systematics, which cannot be properly quantified. We also note that this inflation factor also encompasses the potential correlation between the errors of our spectra.

In summary, our fitting approach ensures convergence to the maximum likelihood, but as systematic errors remain difficult to quantify, the absolute uncertainties reported here do not fully capture these systematics. However, given the homogeneous nature of our analysis, the relative uncertainties across targets remain comparable.

\subsection{Results}
\label{subsec:formosa_win_resu}
We applied \texttt{ForMoSA} to the 43 targets of X-SHYNE using the three different models described in Section \ref{subsec:formosa_models} and considering the four wavelength ranges defined in Section \ref{subsec:formosa_meth}. An example of a fit is shown in Figure \ref{fig:example_best_fit}, where the spectrum of 2MASS~0046 (L0) is overlaid with the synthetic spectra from each model that maximized the likelihood during the inversion. We observe that the quality of the model reproduction improves when each spectral band is considered separately. This confirms the findings of \cite{Petrus24}, who concluded that while models can accurately reproduce observational data locally, they fail to do so on a global wavelength range. We also note that, despite an improvement in the quality of the fit when shorter wavelength ranges are considered, certain spectral features, such as the depth of the potassium line, are not well reproduced. This issue was previously reported by \cite{Petrus23}, who performed dedicated fits to the potassium lines to partially correct the problem. We plan to conduct a similar synthetic analysis, focusing on medium-resolution spectral features, in future work.

\begin{figure*}[h!]
\centering
\includegraphics[width=1.0\hsize]{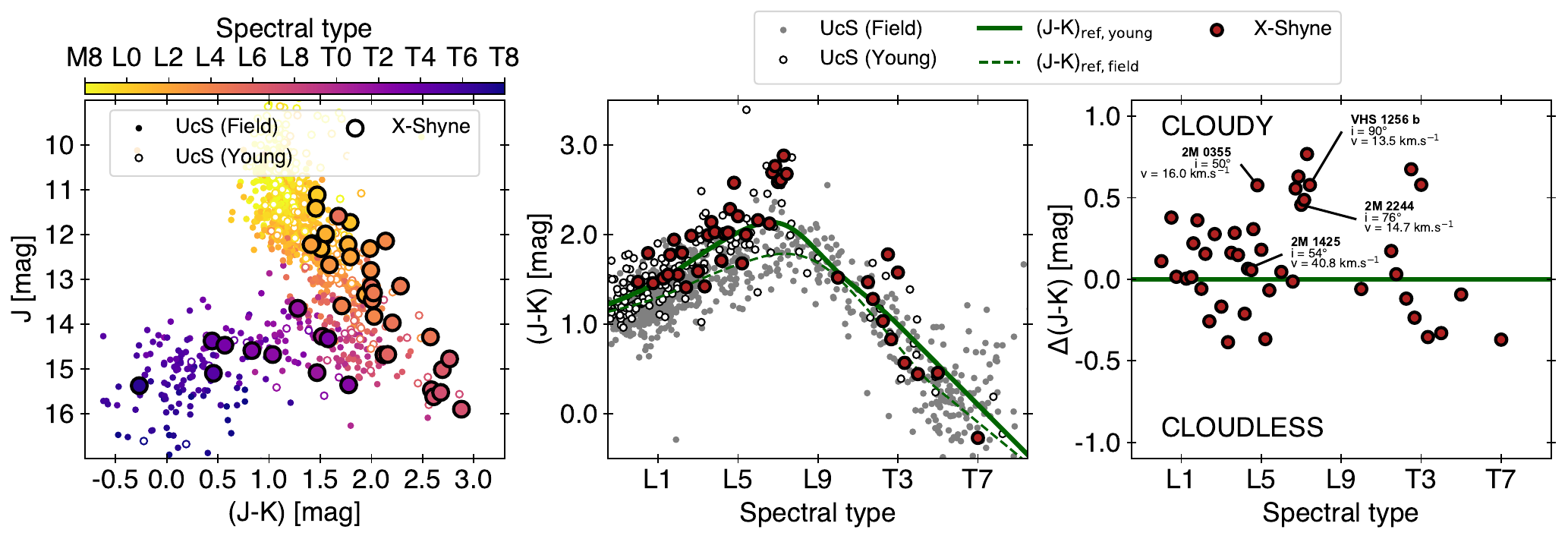}
    \caption{Process used to calculate the photometric anomaly $\Delta$(J-K). Left: Color-magnitude diagram of the L-T transition. We recalculated the J and K magnitudes for all targets from our X-Shooter spectra and compared them with objects from the UltraCoolSheet library. Center: We calculated (J-K) for each object and plotted it as a function of spectral type. The mean values for field and young objects from the UltraCoolSheet library are shown with dashed and solid green lines, respectively. Right: We calculated $\Delta$(J-K) as the offset from the mean (J-K) of young objects with the same spectral type. The four X-SHYNE objects for which rotational velocity and viewing angle measurements are available are highlighted. To improve the clarity of the plot, objects sharing the same spectral type have been evenly distributed within a range corresponding to their spectral type $\pm$1.} 
    \label{fig:phot_ano}
\end{figure*}

\section{Photometric anomaly}
\label{sec:phot_ano}

Since the late 1990s, photometric and spectroscopic observations have pointed to the presence of iron and silicate clouds in the photospheres of L-type brown dwarfs. These clouds help explain why these objects appear redder in color-magnitude diagrams \citep{Tsuji96, Ackerman01, Allard01}. It indicates that clouds predominantly form in early L-type brown dwarfs before settling below the photosphere at the L-T transition as the temperature decreases. A few years later, large photometric time-monitoring surveys detected periodic variations in the infrared light curves of brown dwarfs. These variations were interpreted as inhomogeneities in cloud coverage that evolve with the object's rotation \citep{Radigan12, Apai13, Zhou16}. Such analyses allow for the estimation of the rotational period (P). In combination with the projected rotational velocity (v~sin(i)) derived from the broadening of atomic absorption lines detected via high-resolution spectroscopy, and the radius that can be inferred from models of evolution, it can be used to infer the viewing angle (i) of the observed targets. More recently, clouds have been directly detected through the exploration of mid-infrared wavelengths using the Spitzer Space Telescope and now the JWST. These observations revealed a broad silicate absorption feature around 10~\mic \citep{Suarez22, Miles23}. By combining these silicate detections with photometric anomalies and viewing angles, \citep{Suarez23} highlighted a degeneracy among these parameters. This can be explained by the latitudinal dependence of dust cloud density, which is enhanced at the equator due to the object's rotation. This observational result confirms the predictions made by global circulation models, which indicate that cloud settling at the equator is enhanced with increasing rotational velocity \citep{Tan21}.

Consequently, brown dwarfs with similar fundamental properties (e.g., age, mass, \Teff) but viewed at different inclinations or with different rotational velocities could exhibit varying atmospheric properties when their spectra are compared with 1D atmospheric models. To account for this effect in our results, we quantified the amount of cloud coverage observed in our data. However, only 4 out of our 43 targets have a known viewing angle and silicate detection at 10~\mic. Therefore, we calculated the $\Delta$(J-K) color anomaly for all objects and used it as a proxy of the cloud density.

 Given that brown dwarfs with these \Teff\ are expected to be variable, we aimed to calculate the color anomaly at the epoch of our observations. To do this, we extracted the J- and K-magnitudes from our X-Shooter data using the J- and K$\rm_{s}$-2MASS filters and the flux-calibrated Vega spectrum to derive $\rm(J-K)_{X-SHYNE}$. These photometric magnitudes were then compared to those in the UltracoolSheet library\footnote{https://zenodo.org/records/13993077} \citep{Best21}. For this comparison, we distinguished between objects showing signs of youth and field objects, as illustrated in the left panel of Figure \ref{fig:phot_ano}. To establish reference $\rm(J-K)$ values for given spectral types of young and field objects, we applied a moving box method with a width of three spectral subtypes. This box was shifted across the spectral type range covered by the UltracoolSheet library, and the mean (J-K) was calculated for objects of each spectral type within the box. A Gaussian filter was then applied to smooth this curve and mitigate noise caused by a lack of objects at certain spectral types. For spectral types later than T3, the scarcity of young objects prevented us from calculating a mean (J-K). In these cases, we extrapolated the reference values by assuming a linear trend. Both the young and field $\rm(J-K)_{ref}$ values are shown in the center panel of Figure \ref{fig:phot_ano}, alongside the (J-K) values derived for X-SHYNE objects. Once again, we confirm the young age of X-SHYNE objects, which aligns well with the reference values of young objects from the UltracoolSheet library. We defined the color anomaly $\Delta$(J-K) as the difference $\rm(J-K)_{X-SHYNE}-(J-K)_{ref,~young}$ and calculated it for each X-SHYNE object.

These results are displayed in the right panel of Figure \ref{fig:phot_ano}. A significant diversity in $\Delta$(J-K) values is observed. One possible interpretation of this diversity lies in the variation of viewing angles and rotational velocities, two parameters that are presumed to be randomly distributed for directly imaged objects and remain independent of the observational epoch for isolated brown dwarfs and widely separated planetary-mass companions.

We highlighted brown dwarfs for which estimates of the viewing angle and projected rotational velocity are available. Notably, 2MASS~0355 and 2MASS~1425 share similar spectral types (L3-L4) and viewing angles (50$\rm_{-2}^{+2}$° \citealt{Suarez23} and 54$\rm_{-15}^{+36}$° \citealt{Vos20}, respectively) but exhibit different $\Delta$(J-K) values. This discrepancy is consistent with their differing rotational velocities. For 2MASS~1425, its high rotational velocity (40.8~km.s$^{-1}$) likely facilitates more efficient equatorial migration of clouds, in contrast to the slower-rotating 2MASS~0355. Similarly, VHS~1256~b and 2MASS~2244 have comparable spectral types (L7) and rotational velocities (13.5 and 14.7~km.s$^{-1}$, respectively), but VHS~1256~b shows a higher $\Delta$(J-K) probably due to its equator-on viewing angle \citep{Zhou20}, whereas 2MASS~2244 is observed at a lower inclination (76$\rm_{-20}^{+14}$° \citealt{Vos18}). These trends seem to confirm a link between cloud density, viewing angle, and rotational velocity. However, the radius considered to estimate the inclination is model-dependent. Therefore, These results need to be confirmed by estimating these parameters for the remaining objects in the X-SHYNE sample following a homogeneous analysis. Moreover, the photometric anomaly could be influenced by the high variability of the objects, suggesting a variable cloud density along the longitudinal axis as well. For the remainder of this paper, we simplify the analysis by defining two regimes to qualitatively describe cloud density: $\Delta$(J-K)~>~0, corresponding to a cloudy viewing surface, and $\Delta$(J-K)~<~0, corresponding to a cloudless viewing surface.

\section{Discussion}
\label{sec:discussion}
In Sections \ref{sec:semiempirical_analysis} and \ref{sec:syntetic_analysis}, we estimated the properties of the brown dwarfs in the X-SHYNE spectral library using the predictions of evolutionary and atmospheric models, respectively. In the case of evolutionary models, clouds are not considered, as the properties are derived from the cooling of brown dwarfs over time, driven by the lack of atomic fusion in their cores. It is now well established that clouds have a significant impact on the spectra of brown dwarfs and represent one of the major challenges for atmospheric models to address. In Section \ref{sec:phot_ano}, we utilized the photometric anomaly, $\Delta$(J-K), as a proxy for the impact of clouds along the line of sight. In this section, we discuss our estimates of the properties of X-SHYNE brown dwarfs derived from both approaches and examine the influence of clouds on these properties.

\subsection{Bolometric luminosity, temperature and radius}
\label{subsec:semi_empirique_formosa}

\begin{figure*}[h!]
\centering
\includegraphics[width=0.9\hsize]{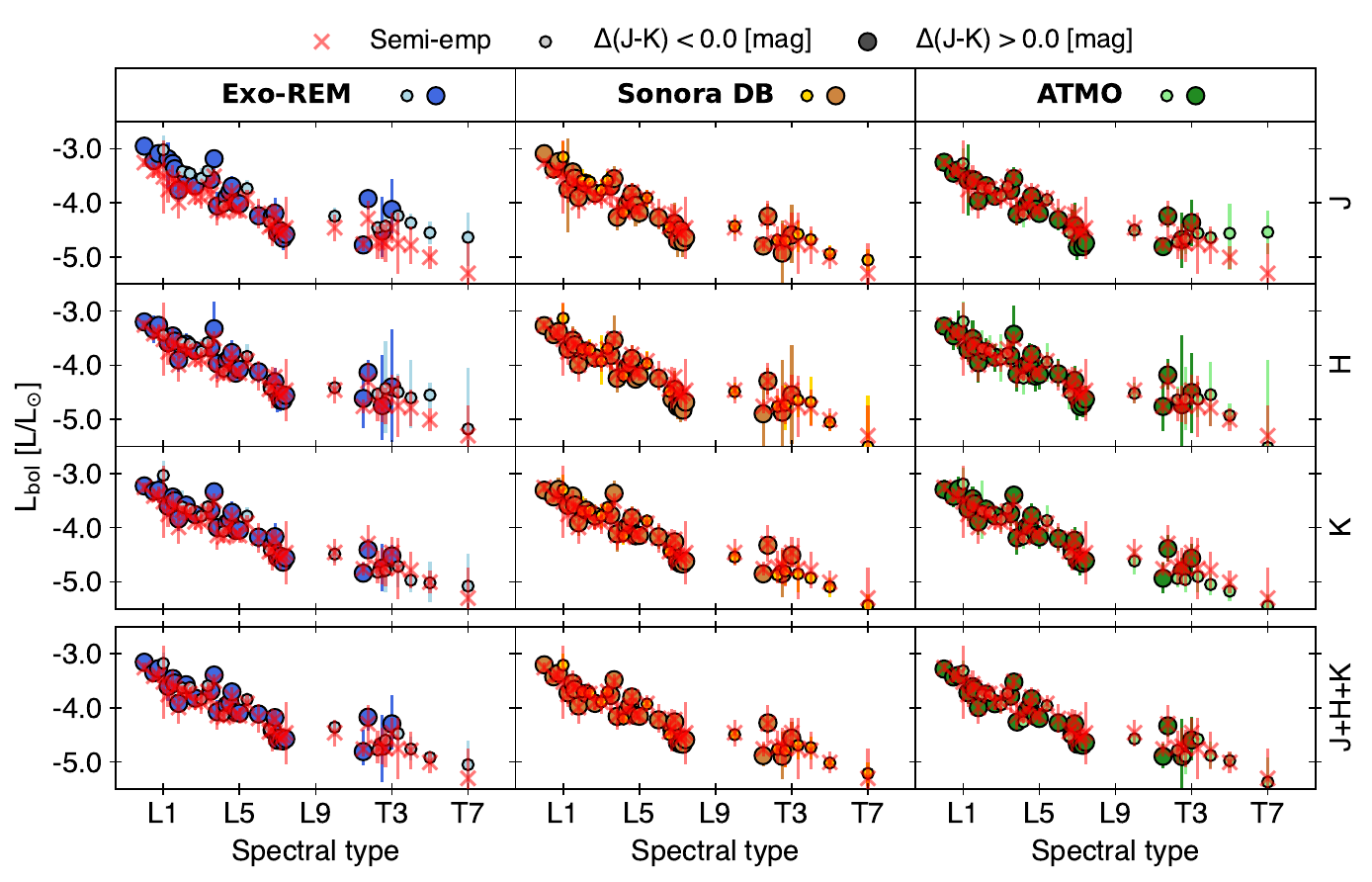}
    \caption{\Lbol estimated from each model grid and for each spectral range using the Stefan-Boltzmann law, based on the estimated \Teff (see Figure \ref{fig:tem}) and radius (see Figure \ref{fig:rad}). Objects with a negative color anomaly are represented by small light dots, while those with a positive color anomaly are shown as large deep dots. The \Lbol values derived from the reconstructed SED (see Section \ref{sec:semiempirical_Lbol}) are indicated by red crosses.}
    \label{fig:lum}
\end{figure*}

\begin{figure*}[h!]
\centering
\includegraphics[width=0.9\hsize]{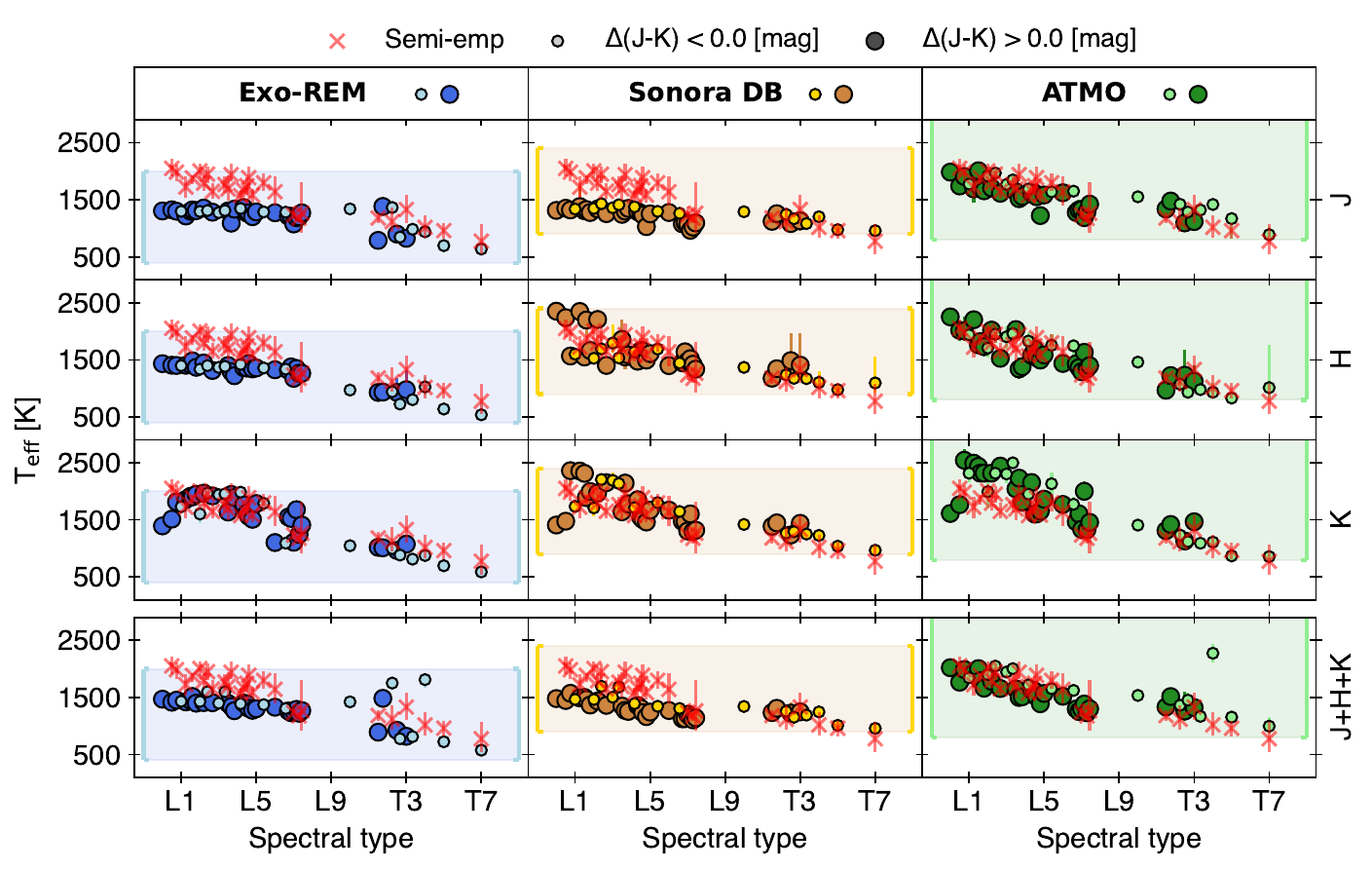}
    \caption{Same as in Figure \ref{fig:lum}, but for the \Teff. The boundaries of each model grid are represented by the filled colored regions.}
    \label{fig:tem}
\end{figure*}

\begin{figure*}[h!]
\centering
\includegraphics[width=0.9\hsize]{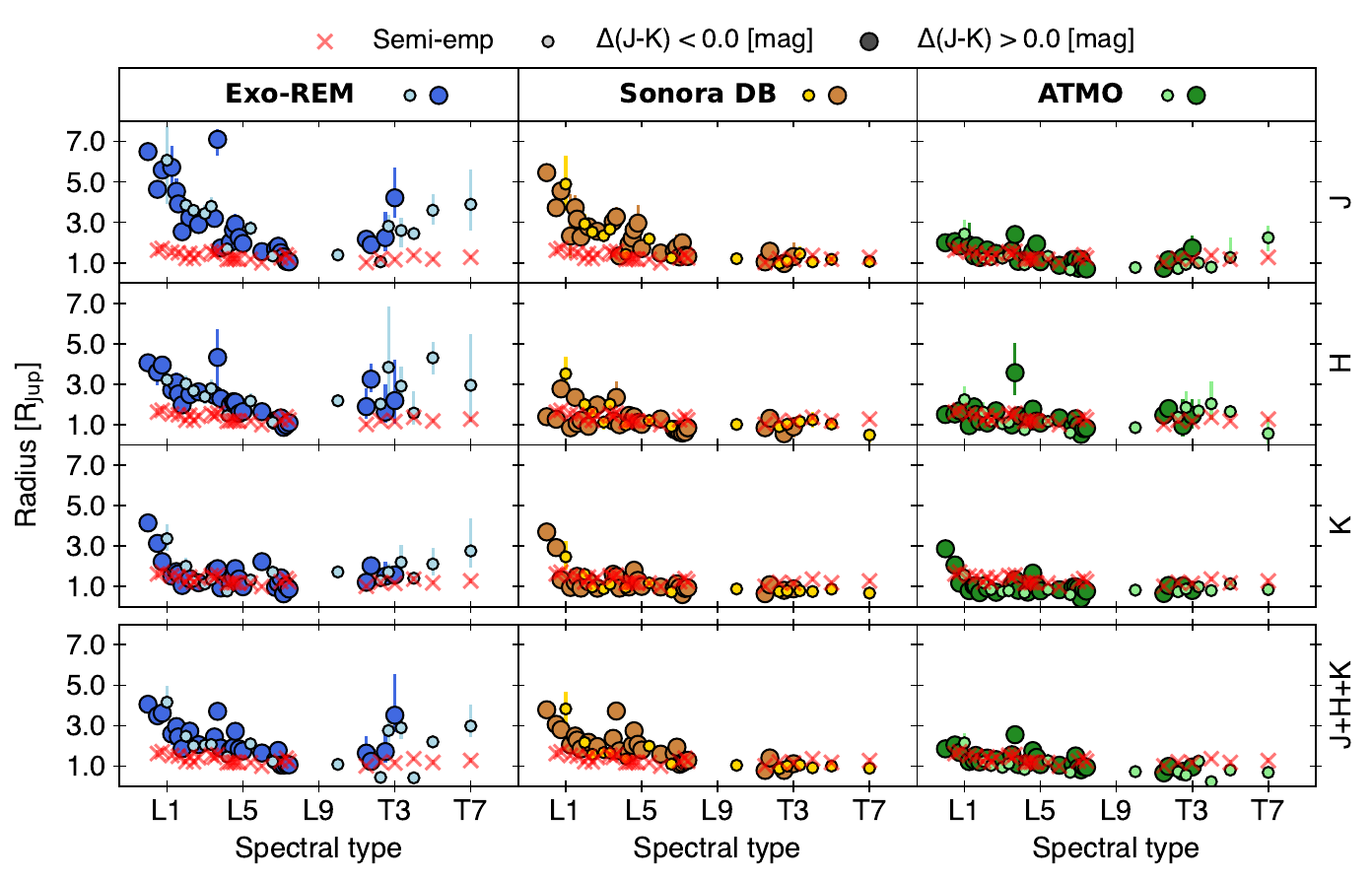}
    \caption{Same as in Figure \ref{fig:lum}, but for the radius.}
    \label{fig:rad}
\end{figure*}

Figures \ref{fig:lum}, \ref{fig:tem}, and \ref{fig:rad} present the estimated values of \Lbol, \Teff, and radius, respectively, for each object in X-SHYNE. The \Teff is derived from model grids, while the radius is inferred from the dilution factor $\rm C_{K}$. The luminosity is calculated using the Stefan-Boltzmann law based on the \Teff and radius estimates, meaning it is not a free parameter in the Bayesian inversion. In each figure, the parameter values predicted by the evolutionary model (see Section \ref{sec:semiempirical_analysis}) are represented by red crosses for direct comparison. For each model, the \texttt{ForMoSA} results distinguish between objects with a positive color anomaly (red excess) and those with a negative color anomaly (blue excess).

Figure \ref{fig:lum} shows that, across all models and spectral bands, the luminosity derived from the synthetic analysis is consistent with the results described in Section \ref{sec:semiempirical_Lbol}. This confirms that self-consistent models can reliably estimate a target’s bolometric luminosity, even when only a limited portion of the SED is available, as demonstrated in previous studies (e.g., \citealt{Carter23}).

In contrast, Figure \ref{fig:tem} highlights the limitations of the \texttt{Exo-REM} and \texttt{Sonora~Diamondback} models in constraining \Teff for brown dwarfs earlier than L6, particularly when using the J or H bands, despite their grids extending to higher \Teff values. In \cite{Charnay18}, which presents the \texttt{Exo-REM} model, it is stated that the absence of certain condensates, such as Al$_{2}$O$_{3}$, or other absorbing species (e.g., ions) that significantly affect the atmosphere at higher \Teff ($>$1500~K), could lead to overly blue synthetic spectra. During spectral inversion, these biased synthetic spectra in the grid prevent likelihood maximization, even if their corresponding \Teff matches the evolutionary model predictions. Consequently, lower \Teff values around $\sim$1500~K are favored, explaining the results in Figure \ref{fig:tem}. This result is in line with the study of \cite{2023ApJ...959...63S} who also found a significant discrepancy between BT-SETTL atmospheric model and evolutionary models at the M-L transition due to a lack of absorbing species.

For \texttt{Sonora~Diamondback}, a similar explanation applies, although the missing condensates may differ. Since Al$_{2}$O$_{3}$ clouds are included in this model, the \Teff values derived from H-band fits are more consistent with semi-empirical estimates. However, when using the K band, both \texttt{Exo-REM} and \texttt{Sonora~Diamondback} yield \Teff estimates that align with evolutionary model predictions, indicating a reduced impact of missing species at these wavelengths.

For \texttt{ATMO}, the derived \Teff values agree with semi-empirical results across all fit configurations except in the K band, where they appear overestimated. Unlike cloud-based models, \texttt{ATMO} replicates cloud effects through atmospheric instabilities (\citealt{Tremblin15}, \citealt{Tremblin16}). This alternative approach eliminates the issue of missing condensates, leading the spectral inversion to converge toward higher \Teff values.

The biases in \Teff observed for each model directly affect the estimated radius, which adjusts during inversion to maintain consistency with the bolometric luminosity of each object. Specifically, when \Teff is underestimated, the radius is overestimated, as shown in Figure \ref{fig:rad}. 

Across all models, a trend emerges where objects with lower $\Delta$(J-K) values appear hotter than those with higher $\Delta$(J-K) values, particularly in the J band. If the color anomaly correlates with cloud density, this \Teff gradient could be explained by cloud absorption, which cools the upper atmospheric layers. In cases of negative color anomalies, low cloud densities allow deeper atmospheric layers, expected to be warmer, to become observable. However, this trend is not seen in T dwarfs, which is unsurprising since these cooler objects are expected to be cloud-free.

\subsection{Cloud sedimentation and adiabatic index}

\begin{figure*}[h!]
\centering
\includegraphics[width=0.9\hsize]{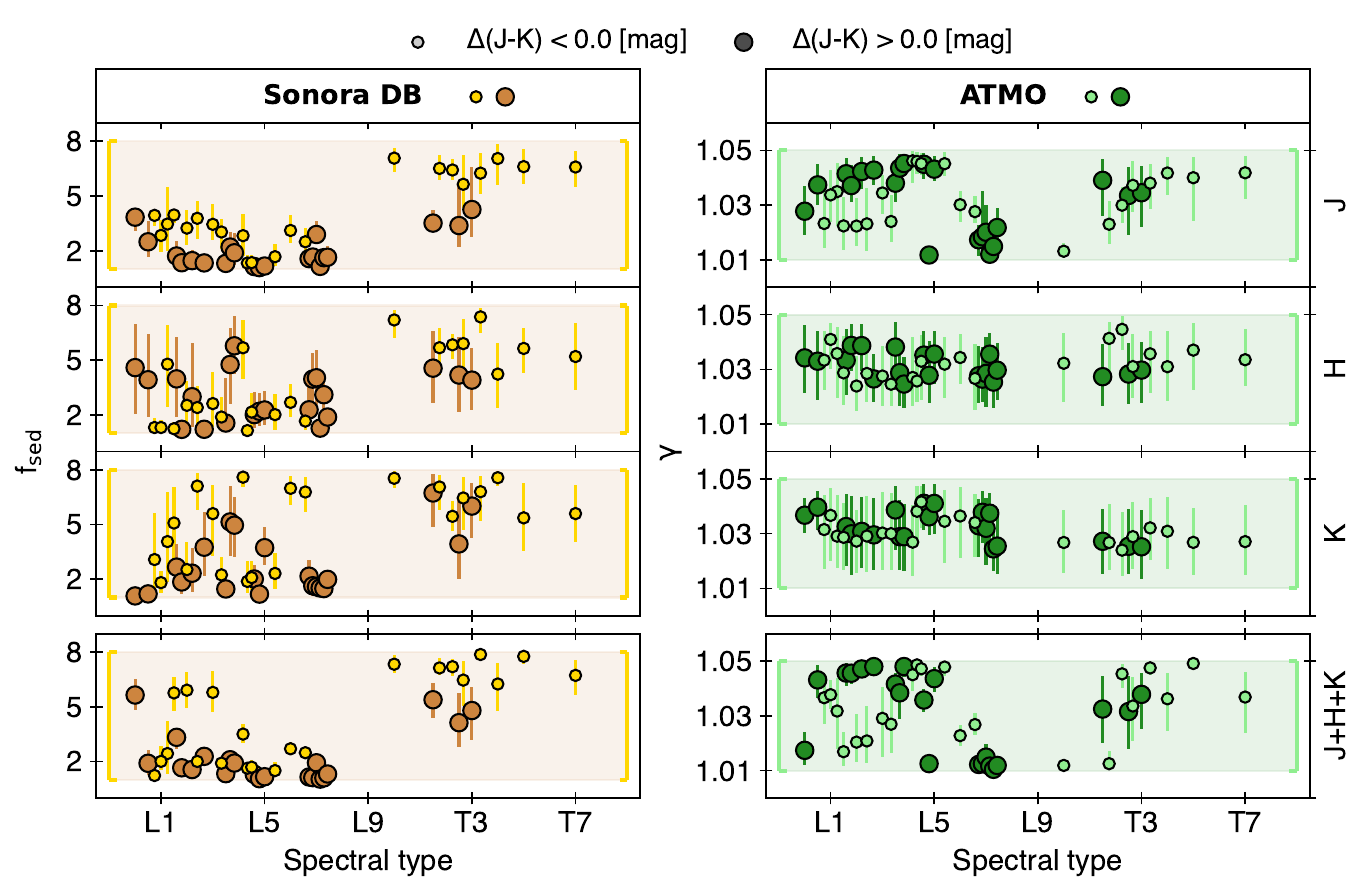}
    \caption{Same as in Figure \ref{fig:lum}, but for the \fsed (left) and the $\gamma$ (right). We notice that because these parameters are not considered in the evolutionary models, the red crosses are not present.}
    \label{fig:fsed_vs_gamma_vs_kzz}
\end{figure*}

To investigate the density of clouds observed in our sample, \texttt{Sonora~Diamondback} allows us to explore different sedimentation efficiencies through its parameter \fsed, while \texttt{ATMO} provides a range of adiabatic index ($\gamma$) values that regulate the strength of atmospheric instabilities. In Figure \ref{fig:fsed_vs_gamma_vs_kzz}, we plotted the estimated \fsed, and $\gamma$ values for each object in X-SHYNE considering the different spectral bands.

As expected, before the L-T transition, we observe low \fsed values ($<$~2), indicating a cloudy atmosphere, whereas after the L-T transition, higher \fsed values ($>$~4) dominate, corresponding to lower cloud densities. Notably, bluer objects tend to have higher \fsed values compared to redder ones, supporting a correlation between this parameter and the color anomaly. Indeed, a lower value of \fsed means a cloudier atmosphere, and therefore a redder spectrum.

The trend for $\gamma$ is less distinct. When considering the J-band and the full wavelength range, we find high $\gamma$ values ($>$~1.03) before the L-T transition, indicating strong atmospheric instabilities, while post-transition values are more moderate ($\sim$~1.03). However, this pattern is less pronounced in the H and K bands. Similar to \fsed, we generally observe lower $\gamma$ values for bluer objects. This is expected since atmospheric instabilities directly contribute to the reddening of the spectra.

\subsection{Surface gravity}

\begin{figure*}[h!]
\centering
\includegraphics[width=0.9\hsize]{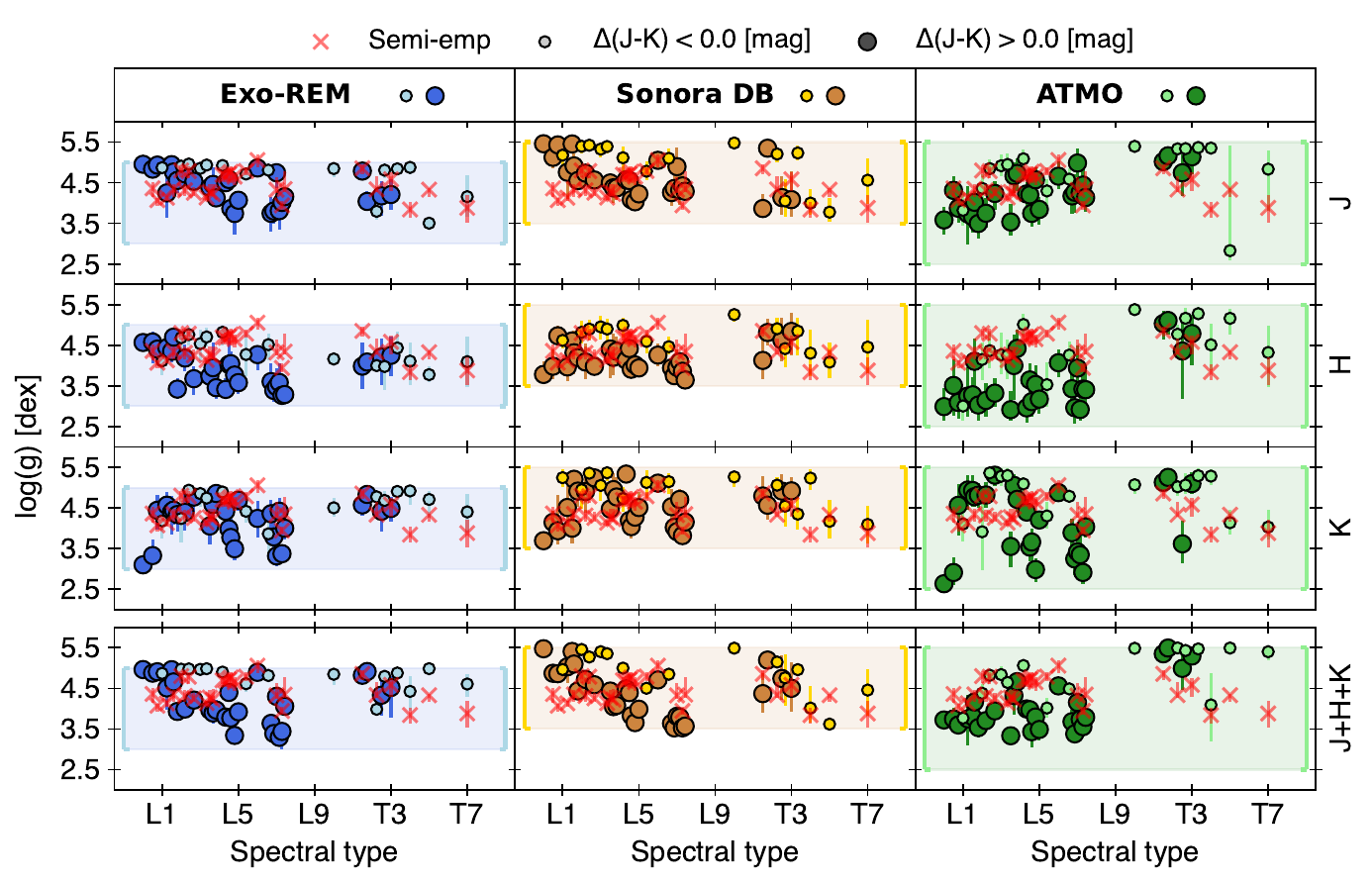}
    \caption{Same as in Figure \ref{fig:lum}, but for the log(g).}
    \label{fig:gra}
\end{figure*}

All the models used here allow us to estimate log(g). In Figure \ref{fig:gra}, we observe that none of these models provide results fully consistent with the predictions of the evolutionary models. Moreover, these results are model-dependent, limiting our ability to determine a robust value for surface gravity. This issue has already been noted in previous multi-model analyses of single objects, such as \cite{Palma23} and \cite{Petrus24}. With X-SHYNE, we are able to capture and visualize these discrepancies across a wide range of spectral types.

The \texttt{Exo-REM} and \texttt{Sonora} models yield similar results, particularly when the J-band is used for the fit. A clear dependence on $\Delta$(J–K) is observed, with redder objects displaying lower log(g) than bluer ones. Low-gravity photospheres have lower pressure than high-gravity ones, resulting in reduced collision-induced absorption (CIA) by H$_{2}$. Since CIA absorbs light over a broad range from the red end of the H band to most of the K band, low-log(g) models naturally appear redder in J-K. However, a similarly red J-K can also be produced by dust reddening, parameterized through \fsed, making log(g) and \fsed naturally degenerate when explaining $\Delta$(J–K). A similar degeneracy should be detected with [M/H], as higher [M/H] increases absorption, leading to a higher altitude for the photosphere. This results in a lower-pressure environment, similar to the effect of low-log(g), thereby reducing CIA absorption. Consequently, [M/H], dust, and log(g) are all degenerate when interpreting $\Delta$(J–K). We can see this effect of $\Delta$(J–K) on the [M/H] in Figure \ref{fig:mh_vs_co}.

We highlight that since log(g) does not strongly affect spectra beyond this CIA-related impact, it is possible that models primarily determine log(g) based on the residual $\Delta$(J–K) after fitting for dust and [M/H]. This could lead to a wide dispersion in estimated log(g) values, not because log(g) varies significantly across the sample, but because it is the most flexible parameter to accommodate the $\Delta$(J–K). We also notice that the \texttt{ATMO} model does not exhibit a clear correlation between log(g) and $\Delta$(J–K), as the reddening effect is primarily driven by the adiabatic index $\gamma$, which governs atmospheric instabilities.

From the perspective of evolutionary models, surface gravity is calculated based on mass and radius. The latter decreases with age, leading to an increase in log(g). However, this approach does not account for the atmospheric complexities reflected in the diversity of $\Delta$(J–K). Because these two different log(g) do not describe the same physical processes, comparing them may lead to significant discrepancies, which could explain the differences observed in Figure \ref{fig:gra}.

\subsection{Metallicity and carbon-to-oxygen ratio}
\label{subsec:co_vs_mh}

\begin{figure}[h!]
\centering
\includegraphics[width=1.0\hsize]{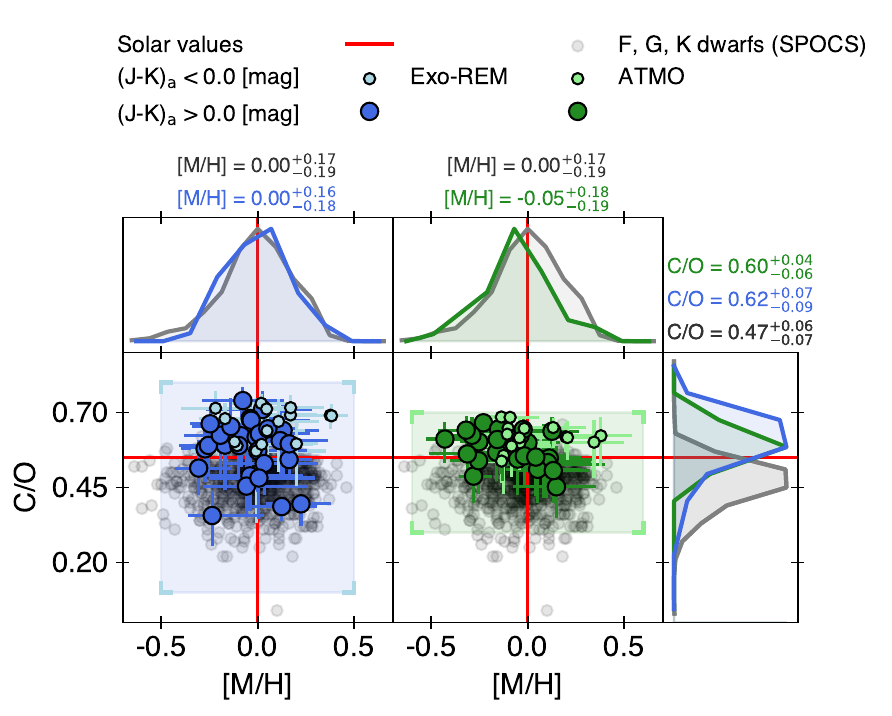}
    \caption{C/O and [M/H] estimated from the \texttt{ATMO} and \texttt{Exo-REM} grids when considering the K band. We overlaid our results for the X-SHYNE library with the abundance estimates from the cool stars in the SPOCS catalog. To facilitate the comparison between both samples, we plotted the distributions of [M/H] and C/O in the right and top panels, respectively. We also provide the mean value and the 1-$\sigma$ standard deviation of each distribution.}
    \label{fig:mh_vs_co_k}
\end{figure}

The relative chemical abundances, C/O and [M/H], are explored in both the \texttt{Exo-REM} and \texttt{ATMO} grids. We compared their estimates when the K-band is considered in Figure \ref{fig:mh_vs_co_k}. This band is expected to be the most sensitive to C/O given the absorptions of CO and H2O in this band. Previous studies have shown that CO overtones at $\sim$2.3~\mic are key features for constraining the C/O ratio (\citealt{Konopacky13}, \citealt{GRAVITY20}, \citealt{Petrus21}, \citealt{Hoch22}, \citealt{Petrus23}). The results for each fitting configuration can be found in Appendix \ref{fig:mh_vs_co}.

From our synthetic analysis, we observe a significant dispersion in the estimated C/O and [M/H] values within the X-SHYNE library. Additionally, the distribution is centered around solar values, suggesting that, overall, the objects of X-SHYNE exhibit solar-like relative abundances.

We also compared our results with the chemical abundances of cool stars from the SPOCS catalog \citep{Valenti05}, which contains properties of 1,040 nearby F, G, and K stars observed through the Keck, Lick, and AAT planet search programs. We used the [M/H] values from \cite{Valenti05} and the C/O estimates from \cite{Brewer16}. Our findings indicate that both \texttt{Exo-REM} and \texttt{ATMO} yield metallicities similar to those of the SPOCS population but suggest slightly higher C/O ratios. Several factors could explain this discrepancy. One possibility is that the X-SHYNE objects formed through different processes than those in the SPOCS catalog, possibly involving enrichment from disk material. However, since our objects are isolated, this scenario is unlikely. Alternatively, systematic errors in the models could introduce biases. Both models assume solar abundances as initial conditions due to the unknown birth environments of the observed objects. However, \cite{Brewer16} show that the C/O measured for the Sun is higher than the C/O measured for the stars from the SPOCS sample. Moreover, spectral data only provide information about the photosphere, meaning the measured abundances may not fully reflect the bulk composition of the objects \citep{Helling14}. Indeed, oxygen can be trapped into clouds and then removed from the photosphere with cloud sedimentation. 

In the context of past and future characterization of imaged planets or young brown dwarfs, the X-SHYNE library offers for the first time a real demographic anchor point for confirming any significant compositional departure that could be linked to heavy element enrichment mechanisms connected to planetary formation processes.  

\section{Summary and conclusion}
\label{sec:conclusion}

In this paper, we exploited the medium resolution (\Rlam$\sim8100$) near-infrared (0.3-2.5~\mic) spectra of 43 young, low-mass brown dwarfs through a semi-empirical and synthetic analysis. With the first one we have calculated the bolometric luminosity of each target by integrating their SED by adapting the method described in \cite{Filippazzo15}. For objects identified as members of young moving groups, we injected the \Lbol and the age of these associations to a model of evolution to estimate the \Teff, the mass, the radius, and the log(g) of our objects. The synthetic analysis has been performed with the code \texttt{ForMoSA} considering the three grids of precomputed synthetic spectra \texttt{Exo-REM}, \texttt{ATMO}, and \texttt{Sonora Diamondback}.

This methodology enabled us to perform a global comparative analysis following three ways of investigation:

\begin{itemize}
   \item Comparison between evolutionary models and atmospheric models: The \Lbol estimates from both approaches are similar. We observed the limitations of the cloudy models, \texttt{Exo-REM} and \texttt{Sonora}, which fail to reach \Teff $>$ 1500~K when the J or H bands are considered due to a lack of absorbers in the simulated atmosphere. In contrast, \texttt{ATMO} provided \Teff values consistent with evolutionary models. Consequently, when \Teff is underestimated, the radius is overestimated. Lastly, the log(g), which represents different physical processes in each type of model, appears inconsistent between the two approaches.

    \item Comparison between the different models of atmosphere: We observe a dispersion in the surface gravity estimates derived from the cloudy models \texttt{Exo-REM} and \texttt{Sonora}, particularly for objects with spectral types ranging from L4 to L7, where clouds are expected to dominate. This dispersion is not present in the cloudless model \texttt{ATMO}. We interpret this as direct evidence of a diversity in cloud densities, driven by variations in rotational velocity, which influence cloud migration toward the equator, and a range of viewing latitudes (inclinations). This interpretation is supported by the calculation of a color anomaly, which depends on cloud density and the sedimentation factor explored in \texttt{Sonora}, where cloudier atmospheres exhibit lower sedimentation efficiency. 

    \item Comparison between the objects of X-SHYNE: We confirmed the decrease of \Lbol with spectral type. Additionally, we observed a lower \fsed for L-type objects, indicating a cloudy atmosphere, and a higher \fsed for T-type objects, consistent with a cloudless atmosphere. The estimated relative chemical abundances, C/O and [M/H], exhibit significant dispersion within the X-SHYNE library, making it difficult to provide robust estimates for individual objects. However, the diversity of the sample allows us to discuss the overall chemical composition of X-SHYNE, which suggests solar values. The metallicity aligns well with measurements from a population of F, G, and K stars, while the C/O ratio derived from X-SHYNE appears higher. This offset could be attributed to a systematic bias related to the initial chemical abundances assumed in the self-consistent atmospheric models or to the difference between the measured C/O, representative of the photosphere, and the bulk C/O of the objects. Finally, the X-SHYNE library provides for the first time a true demographic anchor for the chemical composition of a population of young brown dwarfs, or so-called exoplanet analogs, given their mass overlap with directly imaged planets. This study can confirm any significant compositional departure from a population that is likely formed via stellar mechanisms and that could be related to heavy element enrichment mechanisms associated with planetary formation processes.
\end{itemize}

These results highlight the strength of comparative analysis using homogeneous datasets of a diverse sample of objects to identify model limitations and derive robust conclusions while minimizing over-interpretation. A complementary study should be conducted in the future to confirm the impact of inclination on atmospheric characterization by incorporating high-resolution data to estimate rotational velocity and mid-infrared spectroscopy to quantify silicate cloud density. Additionally, a combined atmospheric model that integrates both cloud formation and atmospheric instabilities should be considered to accurately reproduce the spectra of L dwarfs and derive physically consistent atmospheric parameters.

\section{Data availability}
The numerical values of the equivalent widths estimated for the four potassium lines in Section \ref{sec:semiempirical_age}, the photometric anomaly $\Delta$(J-K) estimated in Section \ref{sec:phot_ano}, and the properties of the objects of X-Shyne obtained with \texttt{ForMoSA} in Section \ref{sec:syntetic_analysis} are only available in electronic form at the CDS via anonymous ftp to cdsarc.u-strasbg.fr (130.79.128.5) or via http://cdsweb.u-strasbg.fr/cgi-bin/qcat?J/A+A/. The spectra presented in the article are also available following the same CDS reference.

\begin{acknowledgements}
We would like to thank the staff of ESO VLT for their support at the telescope at Paranal and La Silla, and the preparation of the observation at Garching. 
This publication made use of the SIMBAD and VizieR database operated at the CDS, Strasbourg, France.  
This work has made use of data from the European Space Agency (ESA) mission Gaia (https://www.cosmos.esa.int/gaia), processed by the Gaia Data Processing and Analysis Consortium (DPAC, https://www.cosmos.esa.int/web/gaia/dpac/consor576u tium). Funding for the DPAC has been provided by national institutions, in particular the institutions participating in the Gaia Multilateral Agreement.  
We acknowledge support in France from the French National Research Agency (ANR) through project grants ANR-14-CE33-0018 and ANR-20-CE31-0012.  
This publication makes use of VOSA, developed under the Spanish Virtual Observatory (\url{https://svo.cab.inta-csic.es}) project funded by MCIN/AEI/10.13039/501100011033/ through grant PID2020-112949GB-I00. VOSA has been partially updated by using funding from the European Union's Horizon 2020 Research and Innovation Programme, under Grant Agreement Nr.~776403 (EXOPLANETS-A).
This work has benefitted from The UltracoolSheet at http://bit.ly/UltracoolSheet, maintained by Will Best, Trent Dupuy, Michael Liu, Aniket Sanghi, Rob Siverd, and Zhoujian Zhang, and developed from compilations by \cite{Dupuy12}, \cite{Dupuy13}, \cite{Deacon14}, \cite{Liu16}, \cite{Best18}, \cite{Best21}, \cite{Sanghi23}, and \cite{Schneider23}.
S.P. is supported by the ANID FONDECYT Postdoctoral program No. 3240145. The authors acknowledge support from ANID -- Millennium Science Initiative Program -- Center Code NCN2024\_001.
J.-S.J. gratefully acknowledges support from FONDECYT grant 1240738 and from the ANID BASAL project FB210003.
G.-D.M. acknowledges the support of the DFG priority program SPP 1992 ``Exploring the Diversity of Extrasolar Planets'' (MA~9185/1), from the European Research Council (ERC) under the Horizon 2020 Framework Program via the ERC Advanced Grant ``ORIGINS'' (PI: Henning), Nr.~832428,
and via the research and innovation programme ``PROTOPLANETS'', grant agreement Nr.~101002188 (PI: Benisty).
Parts of this work have been carried out within the framework of the NCCR PlanetS supported by the Swiss National Science Foundation.
Z.Z. acknowledges the support of the NASA Hubble Fellowship grant HST-HF2-51522.001-A.
A.B. acknowledges the Deutsche Forschungsgemeinschaft's (DFG, German Research Foundation) support under Germany's Excellence Strategy – EXC 2094 – 390783311.
M.C.L. acknowledges the Gordon and Betty Moore Foundation through grant GBMF8550 that funded a part of this work. 
\end{acknowledgements}

\bibliographystyle{aa} 
\bibliography{biblio}

\appendix

\section{Observing log}

Table \ref{Tab:obs_log} presents the observing log for each observed epoch of each object in the X-SHYNE library.

\begin{table*}[t]
\centering
\caption{Obs. log}
\label{Tab:obs_log}
\small
\begin{tabular}{llll||llll}
\hline
Target &  Date        & $\langle$ Seeing $\rangle$ & Airmass & Target &  Date        & $\langle$ Seeing $\rangle$ & Airmass  \\
       & [yyyy-mm-dd] &    ["]                     &         &        & [yyyy-mm-dd] &    ["]                     &          \\ 
\hline
2MASS 0030   & 2018-07-28 & 0.86 & 1.20 & 2MASS 2139   & 2023-06-06 & 0.49 & 1.13 \\
-           & 2018-08-16 & 0.60 & 1.08 & 2MASS 2206   & 2018-06-01 & 0.70 & 1.08 \\
2MASS 0045   & 2018-10-09 & 1.12 & 1.34 & 2MASS 2244   & 2018-07-23 & 0.74 & 1.44 \\
-           & 2018-10-17 & 0.68 & 1.35 & 2MASS 2322   & 2018-06-26 & 0.71 & 1.27 \\
2MASS 0046   & 2019-07-15 & 0.65 & 1.28 & 2MASS 2354   & 2019-10-11 & 0.80 & 1.49 \\
2MASS 0103   & 2019-10-10 & 0.68 & 1.62 & PSO 057      & 2018-10-17 & 1.00 & 1.33 \\
-           & 2019-10-11 & 0.55 & 1.40 & -           & 2018-10-18 & 0.70 & 1.33 \\
2MASS 0153   & 2018-10-08 & 0.59 & 1.39 & -           & 2018-10-22 & 0.74 & 1.32 \\
-           & 2018-10-17 & 0.79 & 1.37 & -           & 2018-10-30 & 0.44 & 1.33 \\
-           & 2018-10-18 & 0.68 & 1.37 & PSO 071      & 2018-10-30 & 0.39 & 1.03 \\
2MASS 0219 b  & 2018-10-21 & 0.93 & 1.16 & -           & 2018-10-30 & 0.34 & 1.04 \\
2MASS 0249 c  & 2019-10-10 & 0.74 & 1.10 & -           & 2018-10-30 & 0.62 & 1.10 \\
-           & 2019-10-10 & 0.88 & 1.20 & -           & 2018-10-31 & 0.63 & 1.08 \\
-           & 2019-10-11 & 0.70 & 1.08 & PSO 318      & 2018-06-20 & 0.64 & 1.14 \\
-           & 2019-10-11 & 1.02 & 1.17 & -           & 2018-06-27 & 0.70 & 1.04 \\
2MASS 0326   & 2018-10-21 & 0.82 & 1.18 & PSO 319      & 2018-05-30 & 0.79 & 1.02 \\
2MASS 0342   & 2018-10-08 & 0.68 & 1.39 & SDSS 1110    & 2018-04-29 & 0.94 & 1.13 \\
2MASS 0355   & 2019-09-10 & 0.76 & 1.25 & -           & 2018-05-27 & 0.67 & 1.12 \\
2MASS 0508   & 2020-01-14 & 0.65 & 1.02 & -           & 2018-05-28 & 0.62 & 1.20 \\
2MASS 0512   & 2018-10-31 & 0.48 & 1.01 & SIMP 0136    & 2018-11-16 & 0.89 & 1.23 \\
2MASS 0518   & 2018-11-27 & 0.68 & 1.25 & ULAS 0047    & 2023-08-01 & 1.12 & 1.32 \\
2MASS 0616   & 2019-11-17 & 0.67 & 1.09 & -           & 2023-08-01 & 0.97 & 1.38 \\
2MASS 0723   & 2020-02-06 & 1.01 & 1.02 & ULAS 1316    & 2023-04-11 & 0.62 & 1.30 \\
2MASS 1021   & 2018-04-28 & 0.60 & 1.09 & -           & 2023-04-11 & 1.00 & 1.31 \\
-           & 2018-05-04 & 0.72 & 1.13 & -           & 2023-03-30 & 1.10 & 1.14 \\
2MASS 1119   & 2019-04-15 & 0.94 & 1.04 & VHS 1256 b    & 2018-05-28 & 0.74 & 1.17 \\
2MASS 1147   & 2018-06-16 & 1.50 & 1.16 & -           & 2018-05-28 & 0.68 & 1.05 \\
2MASS 1148   & 2018-04-29 & 0.53 & 1.14 & WISE 0241    & 2022-12-30 & 0.75 & 1.06 \\
-           & 2018-04-29 & 0.57 & 1.04 & -           & 2023-01-19 & 0.94 & 1.17 \\
2MASS 1207   & 2018-04-28 & 0.56 & 1.09 & -           & 2023-02-03 & 0.86 & 1.43 \\
2MASS 1213   & 2019-04-15 & 0.68 & 1.08 & WISE 0528    & 2018-10-22 & 0.62 & 1.21 \\
2MASS 1425   & 2018-06-19 & 1.56 & 1.14 & -           & 2018-10-26 & 0.58 & 1.21 \\
2MASS 1521   & 2023-04-10 & 0.66 & 1.24 & -           & 2018-10-26 & 0.73 & 1.27 \\
-           & 2023-04-10 & 0.66 & 1.41 & WISE 0819    & 2022-12-16 & 0.86 & 1.20 \\
-           & 2023-05-02 & 0.44 & 1.57 & WISE 2216    & 2019-06-05 & 0.72 & 1.42 \\
2MASS 1826   & 2019-04-18 & 1.08 & 1.08 & -           & 2019-06-16 & 0.97 & 1.41 \\
-           & 2019-04-18 & 1.32 & 1.08 & -           & 2019-07-11 & 0.53 & 1.48 \\
-           & 2019-04-30 & 0.87 & 1.10 & -           & 2019-07-12 & 0.82 & 1.44 \\
2MASS 2104   & 2019-05-03 & 0.59 & 1.20 & -           & 2019-07-15 & 0.68 & 1.45 \\
-           & 2019-07-11 & 0.46 & 1.08 & -           & 2019-07-15 & 0.64 & 1.42 \\
-           & 2019-07-12 & 0.63 & 1.05 &             & 		  &      &      \\

\hline
\hline
\end{tabular}
\end{table*}

\section{Zooms into the X-SHYNE library of spectra}

\begin{figure*}[h!]
\centering
\includegraphics[width=0.92\hsize]{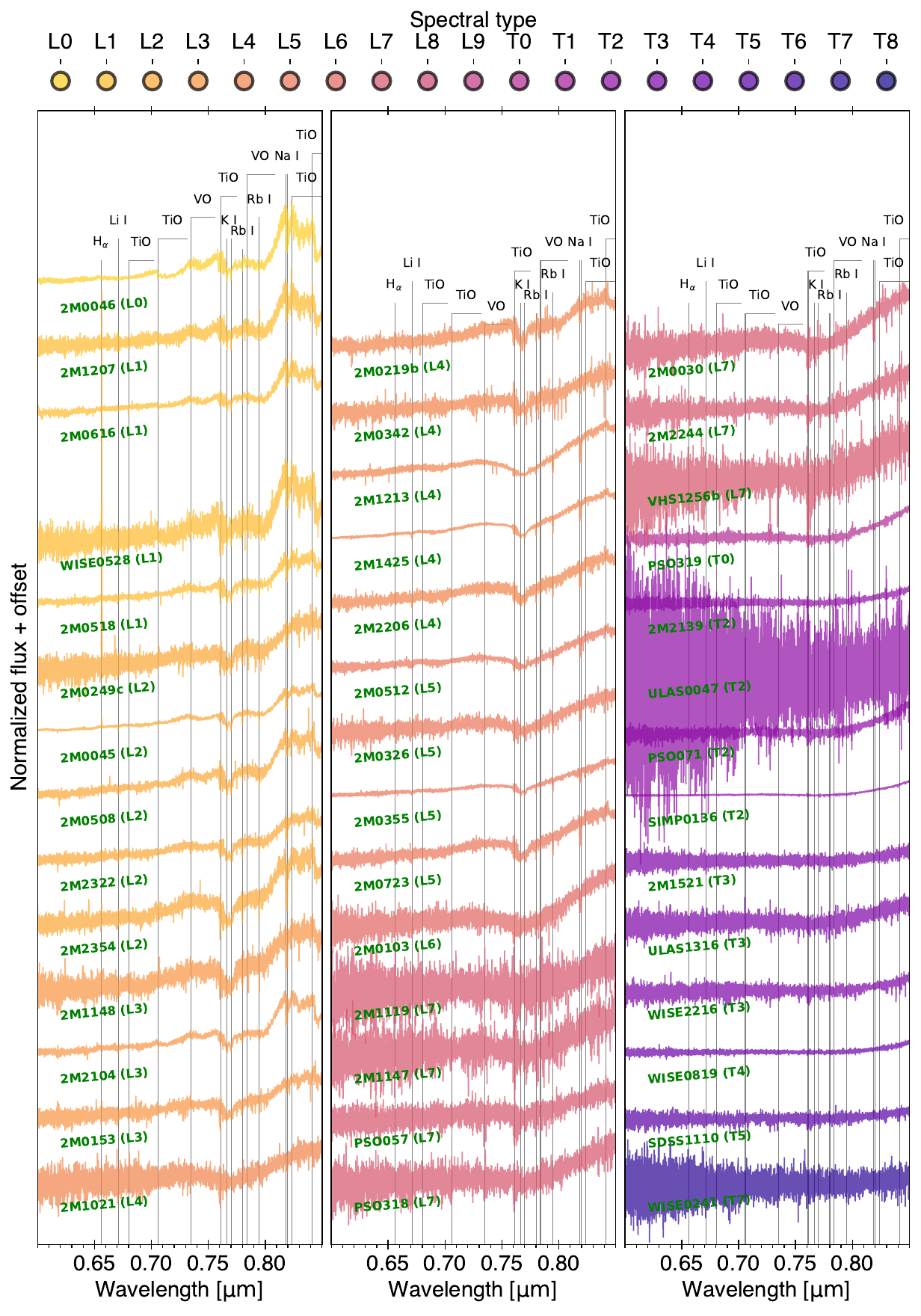}
    \caption{Same than the Figure \ref{fig:sequence_full_wav} with a zoom between 0.60 and 0.85 \mic. The main spectral features are identified.}
    \label{fig:sequence_VIS_wav}
\end{figure*}

\begin{figure*}[h!]
\centering
\includegraphics[width=0.92\hsize]{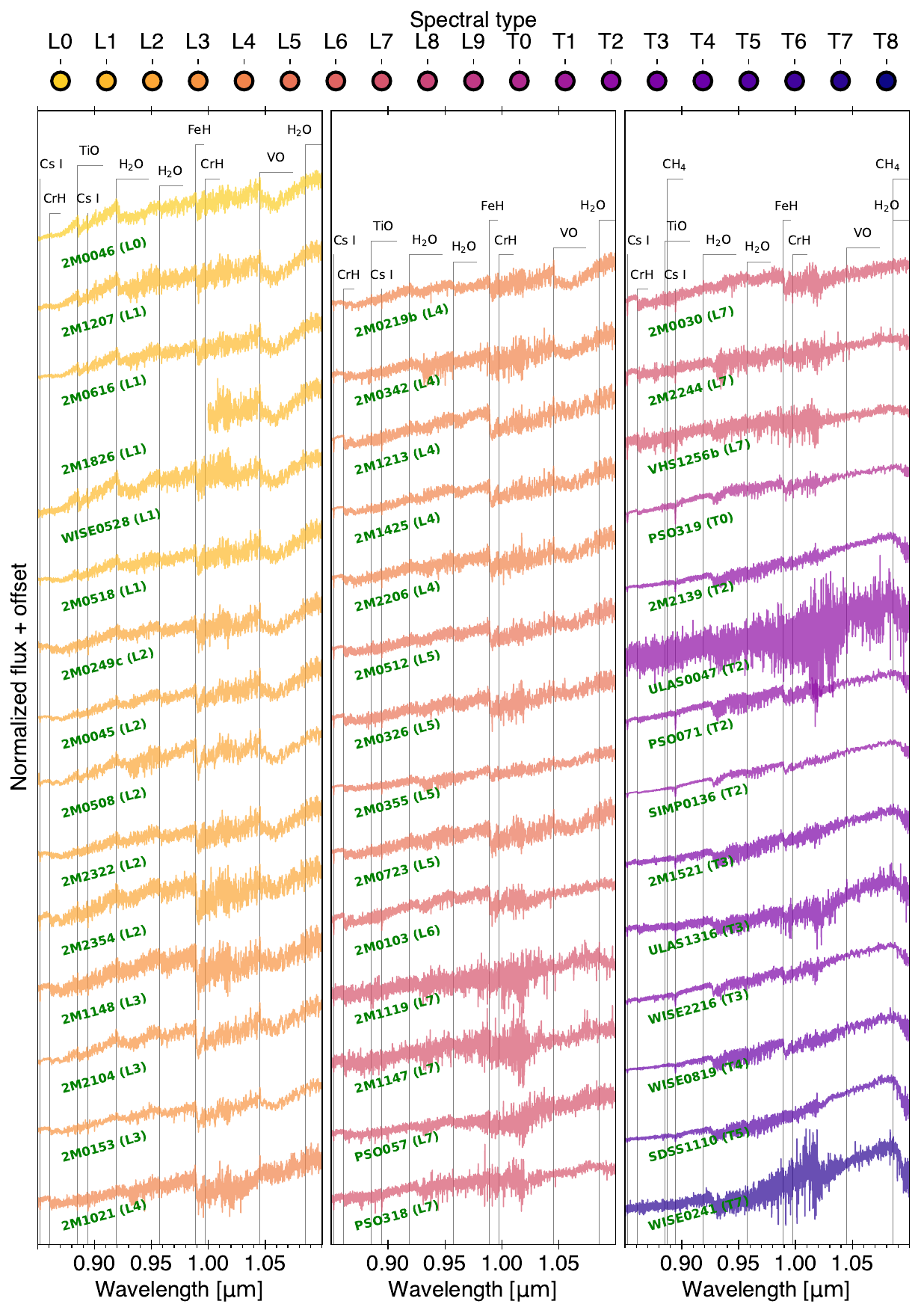}
    \caption{Same than the Figure \ref{fig:sequence_full_wav} with a zoom between 0.85 and 1.10 \mic. The main spectral features are identified.}
    \label{fig:sequence_Y_wav}
\end{figure*}

\begin{figure*}[h!]
\centering
\includegraphics[width=0.92\hsize]{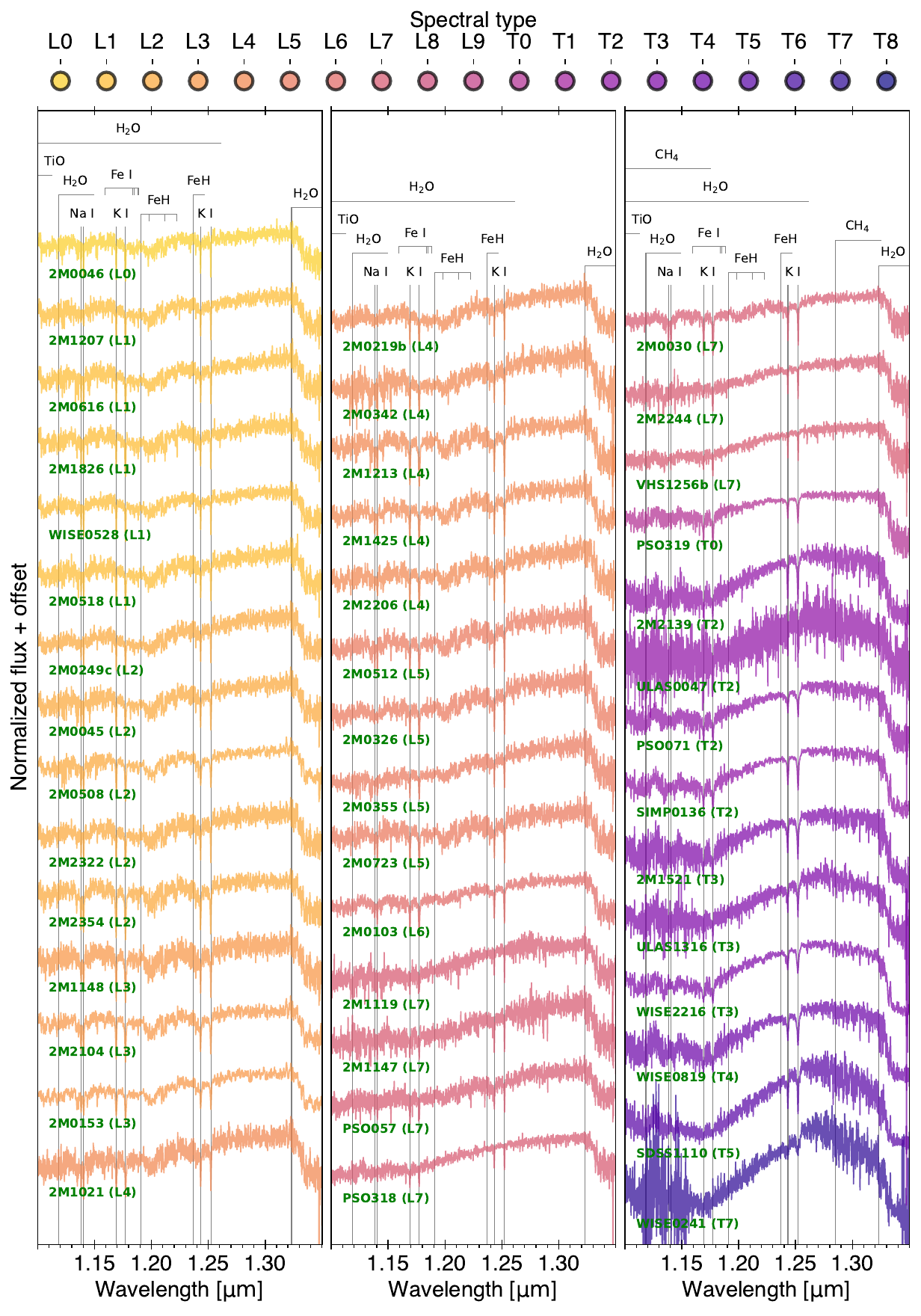}
    \caption{Same than the Figure \ref{fig:sequence_full_wav} with a zoom between 1.10 and 1.35 \mic. The main spectral features are identified.}
    \label{fig:sequence_J_wav}
\end{figure*}

\begin{figure*}[h!]
\centering
\includegraphics[width=0.92\hsize]{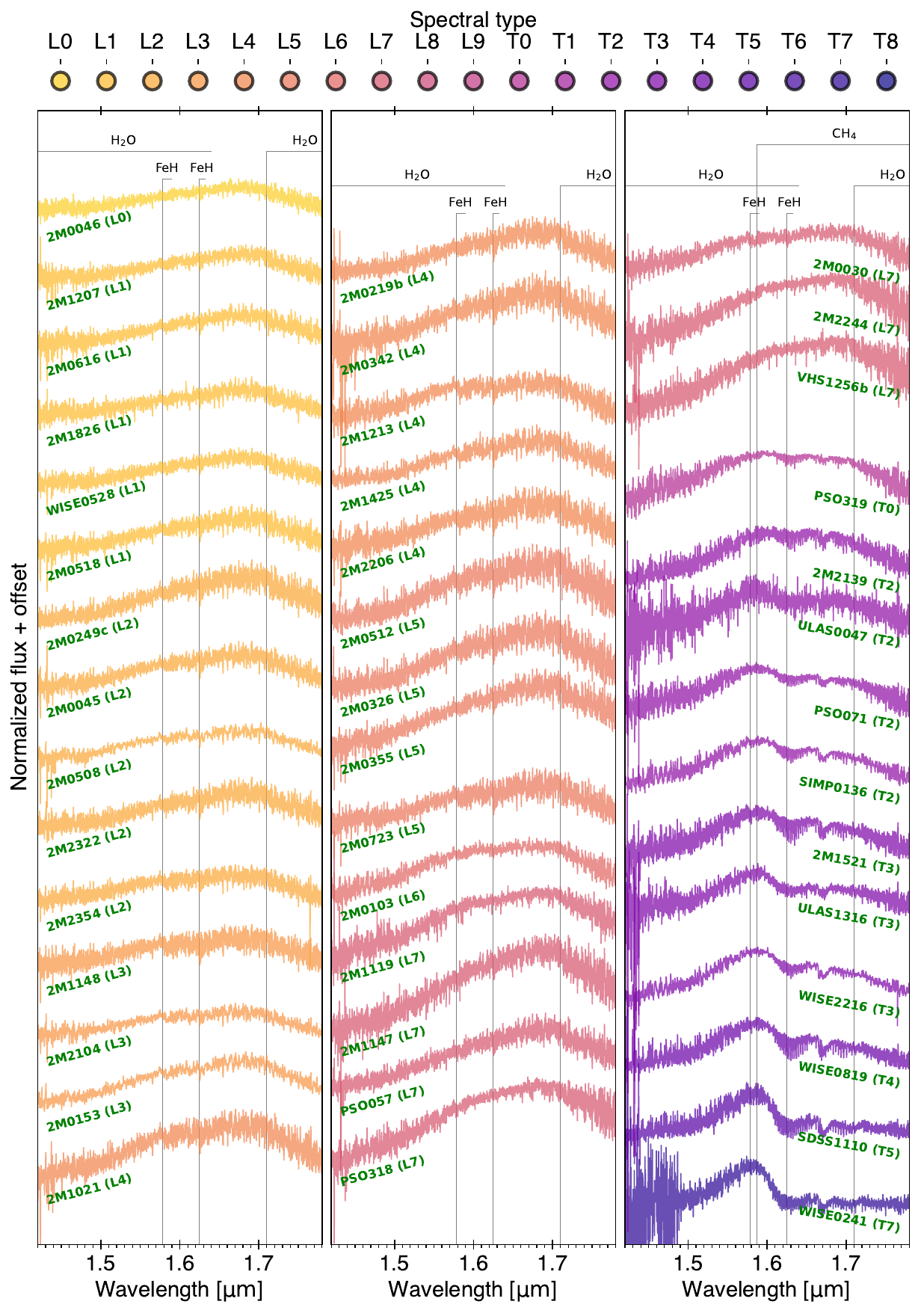}
    \caption{Same than the Figure \ref{fig:sequence_full_wav} with a zoom between 1.42 and 1.81 \mic. The main spectral features are identified.}
    \label{fig:sequence_H_wav}
\end{figure*}

\begin{figure*}[h!]
\centering
\includegraphics[width=0.92\hsize]{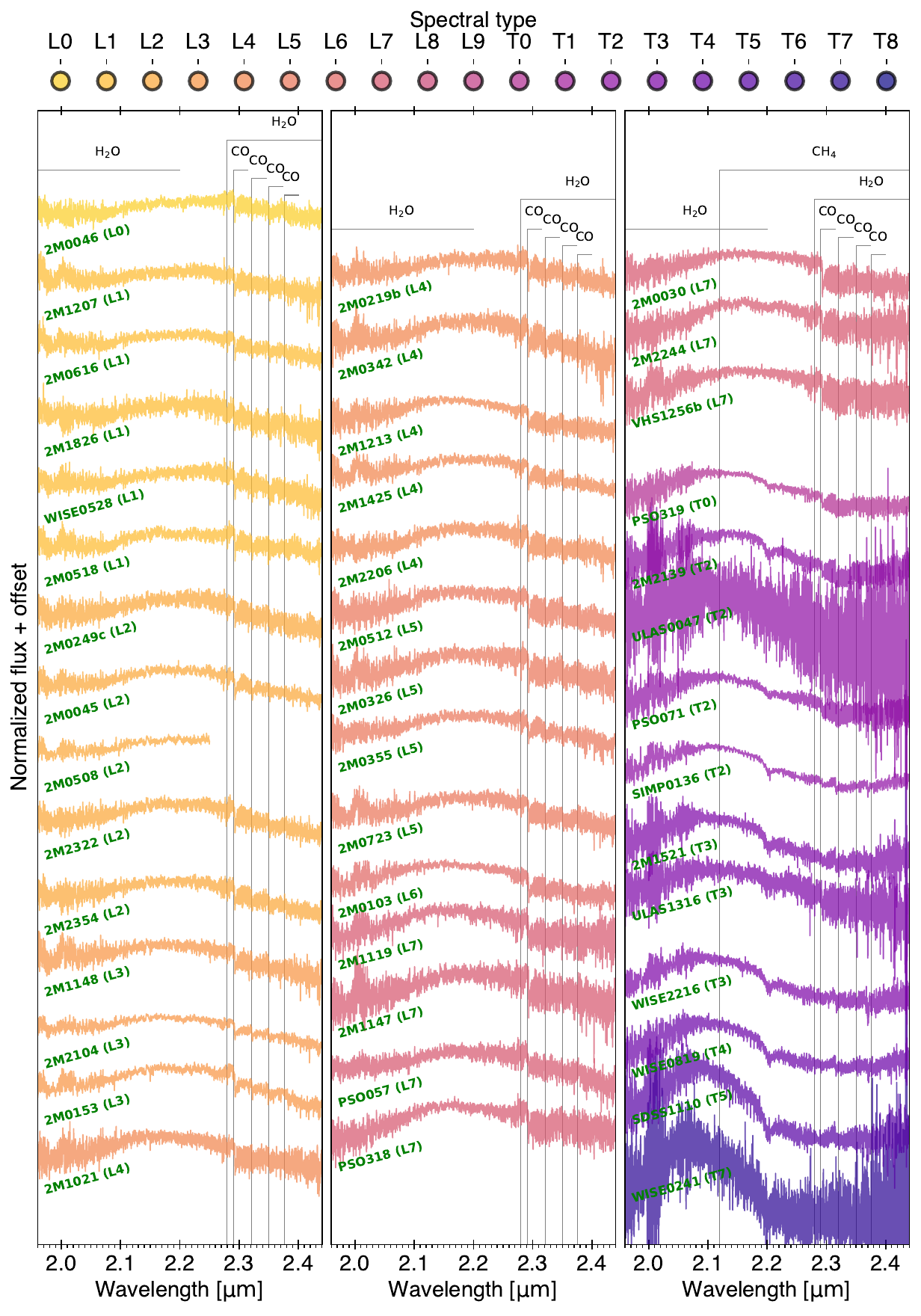}
    \caption{Same than the Figure \ref{fig:sequence_full_wav} with a zoom between 1.96 and 2.48 \mic. The main spectral features are identified. For this plot, the spectra have been normalized with their integrated flux between 1.5 and 1.8 \mic.}
    \label{fig:sequence_K_wav}
\end{figure*}

Figures \ref{fig:sequence_VIS_wav}, \ref{fig:sequence_Y_wav}, \ref{fig:sequence_J_wav}, \ref{fig:sequence_H_wav}, and \ref{fig:sequence_K_wav} illustrate the richness of the X-SHYNE spectra through zoomed-in views of the visible (0.60-0.85~\mic), Y(0.85-1.10~\mic), J (1.10-1.35~\mic), H (1.42-1.81~\mic), and K (1.96-2.48~\mic) bands. Most of the detected atomic and molecular features are highlighted.

\section{[M/H] and C/O ratio for every fitting configurations}

\begin{figure*}[h!]
\centering
\includegraphics[width=1.0\hsize]{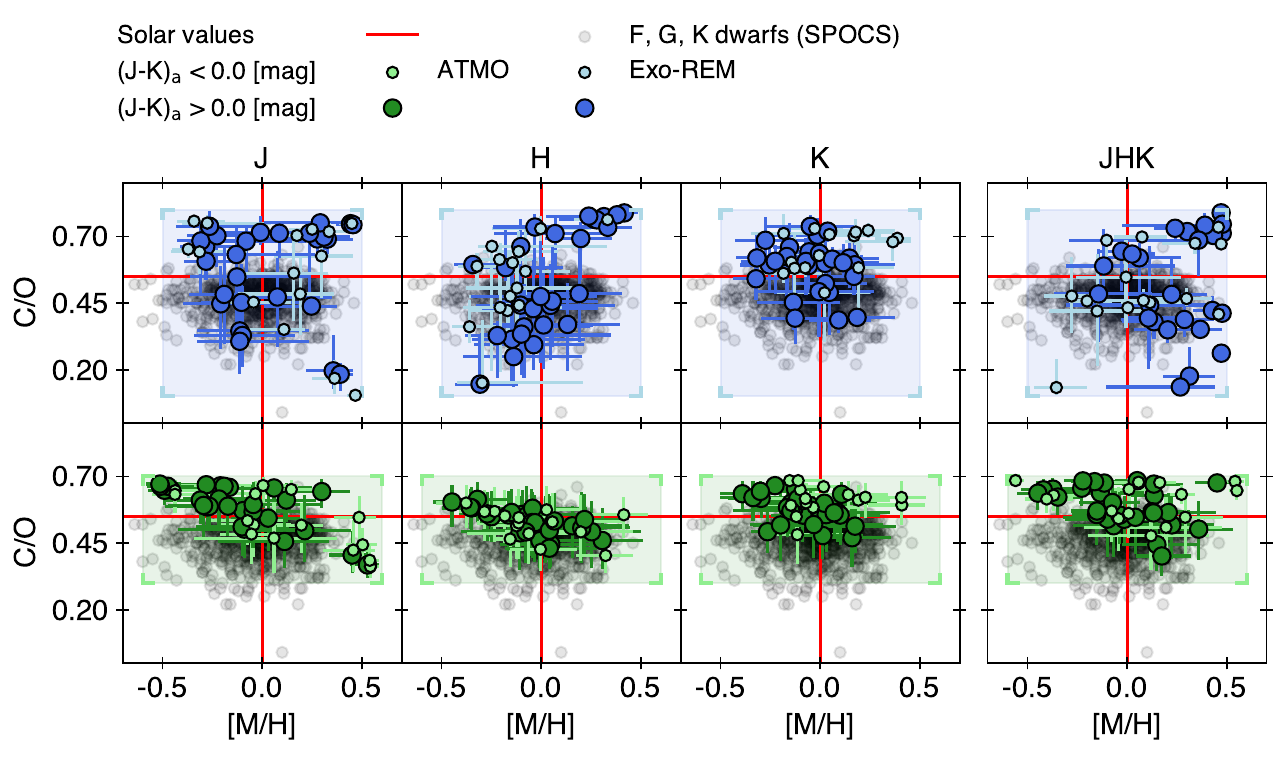}
    \caption{Same as in Figure \ref{fig:mh_vs_co_k} but for each wavelength ranges.}
    \label{fig:mh_vs_co}
\end{figure*}

Figure \ref{fig:mh_vs_co} illustrates the [M/H] and C/O estimates derived from the J, H, and K bands, as well as their combination. The K band provides less dispersed values, likely due to the CO overtones at $\sim$2.3~\mic, which are known to be particularly sensitive to the C/O ratio.

%-----------------------------------------------------------------


\end{document}